\documentclass[sigconf]{acmart}

\usepackage{fontawesome}

\usepackage{xcolor}
\usepackage{soul}
\usepackage{tcolorbox}
\usepackage{wrapfig}
\usepackage{booktabs}
\usepackage{tabularx}
\usepackage[normalem]{ulem}
\usepackage{enumitem}
\usepackage{verbatim}
\usepackage{syntax}
\usepackage[final]{listings}
\usepackage{dialogue}
\usepackage[ruled,vlined]{algorithm2e}

\usepackage[frozencache,cachedir=.]{minted}

\usepackage{tikz}
\usepackage{tikz-qtree}
\usepackage{parcolumns}

\usepackage{tikz}
\usepackage{tikz-qtree}
\usetikzlibrary{trees}
\usepackage{qtree}

\definecolor{key1Color}{HTML}{ca7611}
\definecolor{incrementGreen}{HTML}{669933}
\definecolor{transitionBlue}{HTML}{000033}

\newcommand{\variable}[1]{\textcolor{key1Color}{#1}}

\newcommand{\edited}[1]{\textcolor{black}{#1}}

\newcommand{\intent}[1]{\texttt{#1}}

\newtcbox{\recobox}{nobeforeafter, colback=gray!003, colframe=gray!25, boxrule=0.5pt, arc=1pt, boxsep=0pt,left=2pt,right=2pt,top=1.75pt,bottom=1.5pt,tcbox raise base}

\newcommand{\reco}[1]{\recobox{\small{#1}}}

\newcommand{\snowy}{\textsc{Snowy}}

\AtBeginDocument{%
  \providecommand\BibTeX{{%
    \normalfont B\kern-0.5em{\scshape i\kern-0.25em b}\kern-0.8em\TeX}}}

\copyrightyear{2021}
\acmYear{2021}
\setcopyright{acmlicensed}
\acmConference[UIST '21]{The 34th Annual ACM Symposium on User Interface Software and Technology}{October 10--14, 2021}{Virtual Event, USA}
\acmBooktitle{The 34th Annual ACM Symposium on User Interface Software and Technology (UIST '21), October 10--14, 2021, Virtual Event, USA}
\acmPrice{15.00}
\acmDOI{10.1145/3472749.3474792}
\acmISBN{978-1-4503-8635-7/21/10}

\begin{document}


\title{\snowy: Recommending Utterances for\\ Conversational Visual Analysis}

\author{Arjun Srinivasan}
\affiliation{%
  \institution{Tableau Research}
  \city{Seattle}
  \state{WA}
  \country{USA}
}
\email{arjunsrinivasan@tableau.com}

\author{Vidya Setlur}
\affiliation{%
  \institution{Tableau Research}
  \city{Palo Alto}
  \state{CA}
  \country{USA}
}
\email{vsetlur@tableau.com}

\renewcommand{\shortauthors}{Arjun Srinivasan and Vidya Setlur}


\begin{abstract}
Natural language interfaces (NLIs) have become a prevalent medium for conducting visual data analysis, enabling people with varying levels of analytic experience to ask questions of and interact with their data. While there have been notable improvements with respect to language understanding capabilities in these systems, fundamental user experience and interaction challenges including the \textit{lack of analytic guidance} (i.e., knowing \textit{what} aspects of the data to consider) and \textit{discoverability of natural language input} (i.e., knowing \textit{how} to phrase input utterances) persist. To address these challenges, we investigate utterance recommendations that \emph{contextually} provide analytic guidance by suggesting data features (e.g., attributes, values, trends) while implicitly making users aware of the types of phrasings that an NLI supports. We present \snowy, a prototype system that generates and recommends utterances for visual analysis based on a combination of data interestingness metrics and language pragmatics.
Through a preliminary user study, we found that utterance recommendations in \snowy~support conversational visual analysis by guiding the participants' analytic workflows and making them aware of the system's language interpretation capabilities.
Based on the feedback and observations from the study, we discuss potential implications and considerations for incorporating recommendations in future NLIs for visual analysis.
\end{abstract}



\begin{teaserfigure}
    \centering
    \includegraphics[width=\textwidth,keepaspectratio]{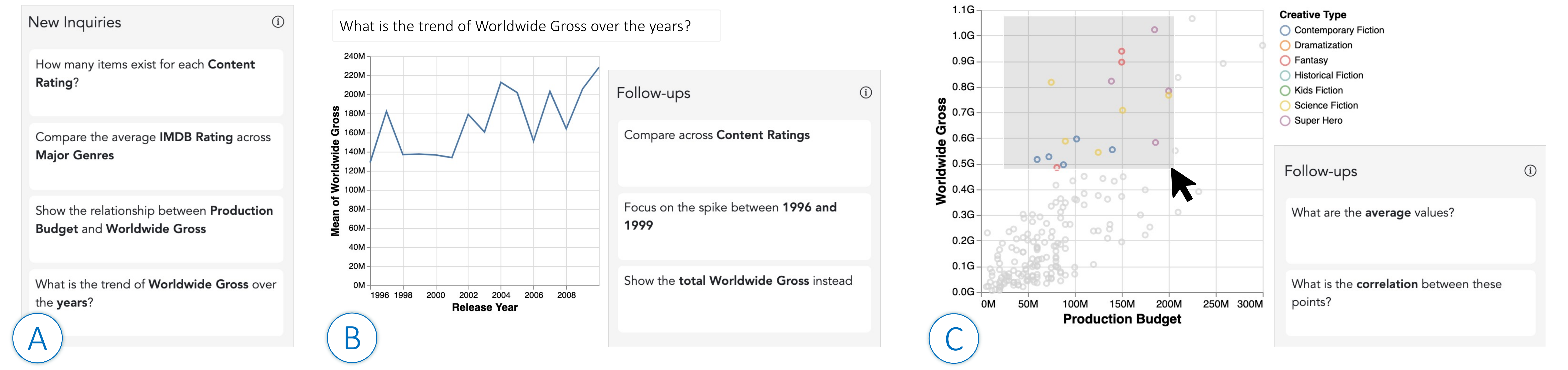}
    \caption{Examples of utterance recommendations in \snowy.
    (A) To assist with the ``cold start'' problem during data analysis, \snowy~infers potentially interesting patterns from the underlying dataset and suggests analytic inquiries one may want to begin exploring the data with.
    (B) Upon executing a NL utterance, \snowy~suggests follow-up utterances to drill down into specific data subsets or adjust the current view.
    (C) As marks are selected on the view through direct manipulation, \snowy~recommends deictic utterances to perform popular calculations using the selected marks.}
    \label{fig:teaser}
    \Description{Figure 1 shows three examples of utterance recommendations in Snowy. Figure 1A shows examples of recommendations at the start of a session. Figure 1B shows recommendations of follow-up utterances presented after processing a user utterance. Figure 1C displays examples of deictic utterances shown when users make selections on a chart.}
    \vspace{1em}
\end{teaserfigure}

\maketitle

\section{Introduction}

Several data visualization tools now support natural language (NL) as an input modality for conducting visual analysis (e.g.,~\cite{sun2010articulate,datatone,setlur2016eviza,narechania2020nl4dv,yu2019flowsense,askdata,powerbi,thoughtspot}).
This interest in NLIs is driven by the fact that NL allows users to freely express their data-driven goals and questions without needing to translate their intentions to interface actions (e.g.~manually specifying mappings between data attributes and visual encodings)~\cite{grammel2010information}.
Recent advances in natural language processing and machine learning have resulted in considerable improvements in NLIs with respect to NL understanding.
NLIs for visual analysis have matured to support a range of analytic intents while also tackling challenges such as ambiguity~\cite{datatone,setlur2016eviza} and underspecification~\cite{setlur2019inferencing}.

Despite improvements in NL understanding, from a user standpoint, formulating utterances during an analytical workflow remains a challenging task for two key reasons.
First, analytical query formulation involves the need to understand the characteristics of the underlying data domain and potential patterns to investigate (e.g., general value distributions, correlations).
A lack of analytic guidance during this process can interfere with users developing an accurate sense of progress towards their analysis goals~\cite{grammel2010information,zenvisage}.
Second, practical limitations of a system's NL understanding capabilities requires users to phrase or adjust their utterances in a way that the underlying system can best interpret them~\cite{setlur2019inferencing} (Here, we use the term \textit{\textbf{utterance}} to refer to any NL command, statement, query, question, or instruction that one may issue to an NLI).
Without a clear understanding of the system's interpretation capabilities, users often end up ``guessing'' utterances, thus being more prone to system failures.
While the \textit{lack of analytic guidance} and \textit{discoverability of NL input} are fundamental challenges on their own, in tandem, these can disrupt the analytic workflow and discourage the use of NLIs for visual analysis altogether.

To address these challenges, we introduce \snowy\footnote{The name Snowy is inspired by the namesake fictional dog in the \textit{Adventures of Tintin} comic series~\cite{tintin} where he addresses his internal monologue to the reader with speech bubbles. He can understand human language and his verbal responses to various situations include jokes, expressions of fright, and pleas to Tintin to exercise caution.}, a novel mixed-initiative interface that presents NL utterances as \textit{recommendations} during visual data analysis (Figure~\ref{fig:teaser}).
\snowy~populates its recommendations with \textit{contextually relevant data entities} (e.g., attributes, values) and phrases the recommendations to highlight the \textit{linguistic variations} supported by the underlying NL understanding module.
This combination of features enables \snowy~to provide analytic guidance that goes beyond existing visualization recommendation tools (e.g.,~\cite{mackinlay2007show,Wongsuphasawat2016,Wongsuphasawat2017,demiralp2017foresight,key2012vizdeck}) that only focus on perceptual features and/or data interestingness, but do not consider NL input, and general NL discovery-focused recommendation tools (e.g.,~\cite{furqan:2017,corbett2016can,voicehints}) that consider the interface and language context, but offer no guidance for visual analysis.

In summary, our key contributions include:
\begin{itemize}
    \item \edited{The design and implementation of} \snowy, a prototype interface that operationalizes the idea of generating and presenting utterance recommendations for conversational visual analysis. \snowy~recommends both 1) \textit{follow-up utterances} that promote a user's active analytic context (e.g., most recent utterance, active chart) as well as 2) utterances that complement the user's historical actions and serve as \textit{new inquiries} to look at other interesting and/or underexplored aspects of the data.
    \item A technique for generating utterance recommendations for visual analysis through a combination of data interestingness metrics (e.g., number of times an attribute has been visualized, correlations between attributes in the underlying dataset) as well as language pragmatics (e.g., terms and entities used in preceding utterances). 
    \item Findings from a preliminary user evaluation of \snowy~suggesting that contextual utterance recommendations can not only guide visual analysis, but also help people gain awareness of the system's NL interpretation capabilities.
\end{itemize}
\section{Related Work}
The primary goal of our work is to support users via recommendations as they use NL as an input modality during their analytical workflows.
Our distillation  of prior research relating to recommendations and NL in the context of visual analysis, falls into three main categories: (1) visualization recommendation tools, (2) NLIs for visual analysis, and (3) user interfaces for NL suggestions.

\subsection{Visualization Recommendation Tools}

Showing visualization recommendations is a popular approach to help users surmount \textit{data selection} and \textit{visual mapping barriers} during visual data analysis~\cite{grammel2010information}.
A detailed review of visualization recommendation (VizRec) systems and techniques is beyond the scope of this paper but can be found in survey manuscripts such as~\cite{lee2021deconstructing,wu2021survey,zhu2020survey,collins2018guidance}.
Broadly speaking, however, VizRec systems can be categorized into 1) systems that focus on recommending visual encodings given a set of data attributes or perceptual constraints (e.g.,~\cite{Mackinlay1986,mackinlay2007show,moritz2019formalizing,wongsuphasawat2016towards}) and 2) systems that leverage recommendations to guide visual data exploration and analysis (e.g.,~\cite{Wongsuphasawat2016,key2012vizdeck,Wongsuphasawat2017,demiralp2017foresight,vismaker:2020}).
We expand upon some tools in the second category since they overlap with the goals of our work.
VizRec systems for data exploration recommend visualizations based on data patterns (e.g.,~\cite{Vartak2015,key2012vizdeck,demiralp2017foresight}) or through a faceted browsing approach showing summaries of attributes in the underlying data (e.g.,~\cite{Wongsuphasawat2016,Wongsuphasawat2017}).
To prune the recommendations and guide analysis, these systems often employ some level of mixed-initiative interaction to steer the system recommendations.
For example, VizDeck~\cite{key2012vizdeck} allows users to rank and organize its recommendations through a voting mechanism.
Systems like Voyager and Voyager 2~\cite{Wongsuphasawat2016,Wongsuphasawat2017} suggest visualizations based on user-selected fields and wildcards to enable rapid iteration through possible data attributes or encodings.
Going beyond attributes, systems like VizAssist~\cite{vizassist} allow users to select their analytic objectives (e.g., finding correlations, creating and comparing data sub-groups) and then create a slew of visualizations with those selections in mind.
Complementing this approach of having users specify objectives, more recent systems like Foresight~\cite{demiralp2017foresight} include predefined categories of ``insights'' (e.g., outliers, dispersion) that are used to recommend visualizations displaying those insights.
We build upon this general space of mixed-initiative tools for visual analysis by considering underlying data patterns and interactions with data attributes over the course of a session as factors to generate recommendations.
However, our work extends this line of research by investigating the idea of generating \emph{utterance} recommendations (as opposed to visualization recommendations).
In doing so, we explore new synergies with research on NLIs for visual analysis and extend the features considered for generating recommendations to not only focus on the underlying data but also concepts from language pragmatics.

\subsection{Natural language interfaces for visual analysis}

NLIs for visualization systems~\cite{askdata,thoughtspot,ibmwatson,powerbi} have evolved over the years to better understand a user's analytical intent expressed in NL and provide reasonable visualization responses.  The forms of inferring intent typically rely on explicitly named data attributes, values, and chart types in the user's input queries. In addition to inferring intent, these systems focus on techniques for enabling users to converse more effectively with such a system. DataTone~\cite{datatone}, for instance, provides ambiguity widgets to allow a user to update the system's default interpretation. Eviza~\cite{setlur2016eviza} and Analyza~\cite{dhamdhere2017} support simple pragmatics in analytical interaction through contextual inferencing. Evizeon~\cite{hoque2017applying} and Orko~\cite{srinivasan2018orko} extend language pragmatics in analytical conversation by understanding follow-up inquiry.
\edited{Iris~\cite{fast2018iris} supports performing complex data science tasks through an NL interface that combines simpler commands through nested conversations.}
Ask Data~\cite{setlur2019inferencing,askdata} handles various analytical expressions in natural language form such as grouping of attributes, aggregations, filters, and sorts. The system also handles impreciseness around numerical vague concepts such as `cheap' and `high' by inferring a range based on the underlying statistical properties of the data. Other research has explored how ambiguity can be handled in NL utterances with reasonable defaults.
Hearst et al.~\cite{hearst2019toward} explore appropriate visualization responses based on the shape of the data distributions for singular and plural superlatives in NL utterances (e.g., `highest price' and `highest prices') and numerical graded adjectives (e.g., `higher'). Sentifiers~\cite{Setlur2020Sentifiers} explores interpretations and defaults for subjective vague modifiers such as `best' and `safe' during visual analysis. 

Despite the advancements in better understanding intent, formulating appropriate NL utterances during data exploration still remains a challenging problem. Users tend to adapt to the suggestions that the system provides even if the system can handle greater degree of variability and underspecificity than what the suggestions provide~\cite{problematicsearch:2020,predictivetext:2020}. The problem is further exasperated as users try to find next steps in their analyses that yield useful insights. In our work, we address this cognitive overload in NLIs by \edited{recommending} utterances with linguistic variation to help guide users during their analytical workflows.
\edited{Furthermore, prior work has also shown that the combination of NL and direct manipulation facilitates an integrated interaction experience and enables higher degree of freedom of expression during visual analysis (e.g.,~\cite{datatone,setlur2016eviza,saktheeswaran2020touch}). Following these findings, we investigate the proposed idea of generating and presenting utterance recommendations in the context of a multimodal visualization interface that supports both manual view specification and NL input.}

\subsection{User interfaces for natural language suggestions}

The lack of input discoverability has been a long standing challenge for NLIs~\cite{walker1998can,yankelovich1996users}.
Given the general nature of this challenge, we referred to prior approaches in this space to identify design challenges and explore potential solutions.
Specifically, we focus on adaptive NL command discovery interfaces that incrementally expose users to the system features through contextual suggestions (e.g.,~\cite{myers2019adaptive,furqan2017learnability,corbett2016can,voicehints}), and explore how such suggestions can be generated during visual analysis.

Query suggestions have also shown to benefit users during exploratory search tasks (e.g.,~\cite{otsuka:2012,medlar:2021}). They are often displayed alongside search results and are intended to be used as follow-on queries or reformulations of the present search query~\cite{bhatia:2011}. Methods for generating query suggestions use information from query logs based on click-throughs or query co-occurrence~\cite{he:2009}. Recent advances support various adaptive techniques to encourage the discovery of new utterances during a user's search journey through visual feedback, in-situ suggestions, and context-sensitive orientation~\cite{adaptivesuggestions:2019}.
\edited{We draw inspiration from these ideas of generating query suggestions based on search relevance and prior user interactions, applying them to the context of conversational visual analysis.}

In the space of visual analysis \edited{tools}, systems like Power BI Q\&A~\cite{powerbi}, Ask Data~\cite{askdata}, and Thoughtspot~\cite{thoughtspot} display textual suggestions as one types a query. SneakPique~\cite{sneakpique} displays widgets with embedded visualizations as visual data previews as users type NL utterances. By doing so, these systems help users rapidly formulate or refine their queries.
However, since they are invoked only when users type \edited{or interact with the text input box}, these interfaces offer little or no assistance in scenarios where users are unaware of what query to begin with (e.g. at the start of data exploration) \edited{or what aspects of the data to consider next during an analytic session}.
Furthermore, the suggestions offered by \edited{current state-of-the-art} tools typically try to showcase the available analytical capabilities and may not be driven by potentially interesting data patterns.
This again imposes the task of figuring out \textit{what} to ask onto the users.
Addressing these gaps, in our work, we provide utterance recommendations that users can utilize at different points of their analyses (e.g., at the start of a session, as a follow-up to a prior utterance, or as a follow-up to a deictic action such as selecting points on a chart). These recommendations not only give users a sense for the linguistic capabilities of the system, but also guide them towards interesting data subsets and patterns to consider.
\section{Recommending Utterances during Visual Analysis}

The key idea driving our research is to explore how \textit{utterance recommendations} (for brevity, we refer to utterance recommendations as \textit{recommendations} hereafter) in NLIs can guide visual analysis while implicitly helping users learn and discover the system's NL understanding capabilities.
Operationalizing this idea, however, requires answering several open questions pertaining to the recommendations as well as the interface within which they are incorporated.
For instance, regarding the recommendations themselves, what information should the recommendations contain (e.g., attribute names, visualization types, keywords mapping to analytic tasks)? How should the recommendations be phrased? Should they be phrased as system commands or more colloquially? Should the recommendations use data attributes and values that users have previously interacted with so they seem more familiar? Or should recommendations promote breadth in interaction and cover aspects of the data that one may not have looked at before?
In terms of their interplay with the interface, when should the recommendations be shown? At the start of a session or during exploration? Should they be generated proactively or be shown on-demand? Which actions should update or trigger recommendations? Since the recommendations show NL utterances, should they only be presented during NL interaction? Or could they also be used to promote multimodal input?

\subsection{Design Goals}
With the above questions in mind, we compiled a list of six design goals to guide \snowy's design. These goals were informed by prior work on mixed-initiative user interfaces~\cite{horvitz1999principles}, exploratory visualization recommendation systems (e.g.,~\cite{Wongsuphasawat2016,Wongsuphasawat2017,demiralp2017foresight,datasite,lee2021deconstructing}), and NLIs for data visualization (e.g.,~\cite{sun2010articulate,datatone,setlur2016eviza,narechania2020nl4dv,hoque2017applying}). We refined the user experiences while iterating over early designs of the prototype.

\vspace{.5em}
\noindent\textbf{DG1. Facilitate breadth in data exploration.}
Prior work on VizRec systems (e.g.,~\cite{Wongsuphasawat2016,demiralp2017foresight,key2012vizdeck}) and exploratory data analysis in general~\cite{tufte1985visual} have stressed on the importance and challenges of breadth-oriented exploration.
Along these lines, to provide effective analytic guidance during data exploration, the recommendations should promote both analytic and data coverage by spanning across a breadth of intents and data entities.

\vspace{.5em}
\noindent\textbf{DG2. Support new inquiry, follow-ups, and deictic utterances.}
Prior work has shown that NL input can be used in different contexts during visual analysis.
Specifically, one-off NL utterances can be used to ask data-driven questions or specify visualizations (e.g.,~\cite{sun2010articulate,datatone,yu2019flowsense,narechania2020nl4dv}).
Alternatively, NL can be used to engage in an analytic dialogue, iteratively create and refine a visualization through a sequence of utterances (e.g.,~\cite{setlur2016eviza,hoque2017applying,setlur2019inferencing}).
Finally, NL can be used in concert with direct manipulation actions like selections to perform multimodal interactions (e.g.,~\cite{setlur2016eviza,srinivasan2020inchorus,srinivasan2020interweaving}).
Hence, the recommendations shown must support these different scenarios of using NL input during visual analysis and include \textit{new inquiry}, \textit{follow-up}, and \textit{deictic} utterances as and when applicable.

\vspace{.5em}
\noindent\textbf{DG3. Use analytic intents as anchors.}
A definitive characteristic that differentiates NL from traditional visualization GUIs is that NL enables people to clearly express their analytic tasks or \textit{intents} (e.g., finding correlations, seeing distribution of values in a dataset).
Thus, the recommendations should map to one or more analytic intents so that users can leverage these intents as anchors to guide their analyses. To cover a spectrum of functionality exhibited by prior NLIs and VizRec systems, we currently focus on six intents: 1) \intent{group}ing data by a categorical attribute (e.g.~``\textit{average sales by region}''), 2) seeing data \intent{distribution} for individual attributes (e.g.~``\textit{Show count of movies by genres}''), 3) understanding the \intent{correlation} between two quantitative attributes (e.g.~``\textit{How does profit vary with sales?}''), 4) observing the \intent{trend} in values over time (e.g.,~``\textit{Display quarterly sales since 2019.}''), 5) \intent{filter}ing to drill down into a specific data subset (e.g.~``\textit{Compare sales for USA, Japan, India, and Australia}''), and 6) changing the \intent{aggregate} function used to calculate derived values (e.g.~``\textit{Show total sales instead of average sales by region}'').

\vspace{.5em}
\noindent\textbf{DG4. Complement user \edited{interaction} with data interestingness.}
Following \textbf{DG1}, the recommendations should consider the users' interactions and suggest underexplored data entities and analytic intents.
However, depending on the size of the dataset, choosing attributes and data subsets to recommend can be difficult and quickly turn into a combinatorial explosion.
Recent ``insight''-based recommendation tools (e.g.,~\cite{datasite,demiralp2017foresight}), on the other hand, have shown that this challenge can be tackled by computationally analyzing interesting patterns in the underlying data and suggesting the most relevant results.
Such recommendations based on patterns in the underlying data can also help with ``cold start'' scenarios where users are new to a dataset and may not have a clear analytical goal in mind.
Thus, the system should consider data interestingness as an additional feature (along with data and analytic coverage) when generating recommendations.

\vspace{.5em}
\noindent\textbf{DG5. Support linguistic variation.}
NL utterances posed to visualization tools often vary from highly specified utterances that contain explicit references to data attributes, chart types, and/or analytic intents (e.g.,~``\textit{Show me a bar chart of average sales by country},'' ``\textit{Show the average profits for countries with over \$50M in sales}'') to highly underspecified utterances that only contain partial references to data entities or intents (e.g., ``\textit{Compare sales around the world},'' ``\textit{Show profits for countries with high sales}'').
While supporting different NL input contexts (\textbf{DG2}), the system must also ensure that the recommendations exhibit ample linguistic variation within and across each context.

\vspace{.5em}
\noindent\textbf{DG6. Should be unobtrusive during targeted analysis.}
While a key goal of the recommendations is supporting open-ended data exploration, there may be instances where users have a clear goal in mind about what they want to look at.
In such scenarios, the recommendations should not interfere with the user's thought process but be available to look at in case users want to refer to the recommendations to identify phrasings that match the query in their mind.
\newline

\noindent{}Note that these goals are not exhaustive or mutually exclusive, nor are they meant to be prescriptive for utterance recommendations in NLIs for data visualization.
For instance, we focus on visual data exploration as our primary usage context and do not consider actions like changing encodings or formatting marks as part of the recommendation space.
Instead, \textbf{DG1}-\textbf{DG6} are only meant to be an initial set of goals to help ground our design and enable us to develop and test an initial prototype in this space.
\begin{figure*}[t!]
    \centering
    \includegraphics[width=.9\textwidth,keepaspectratio]{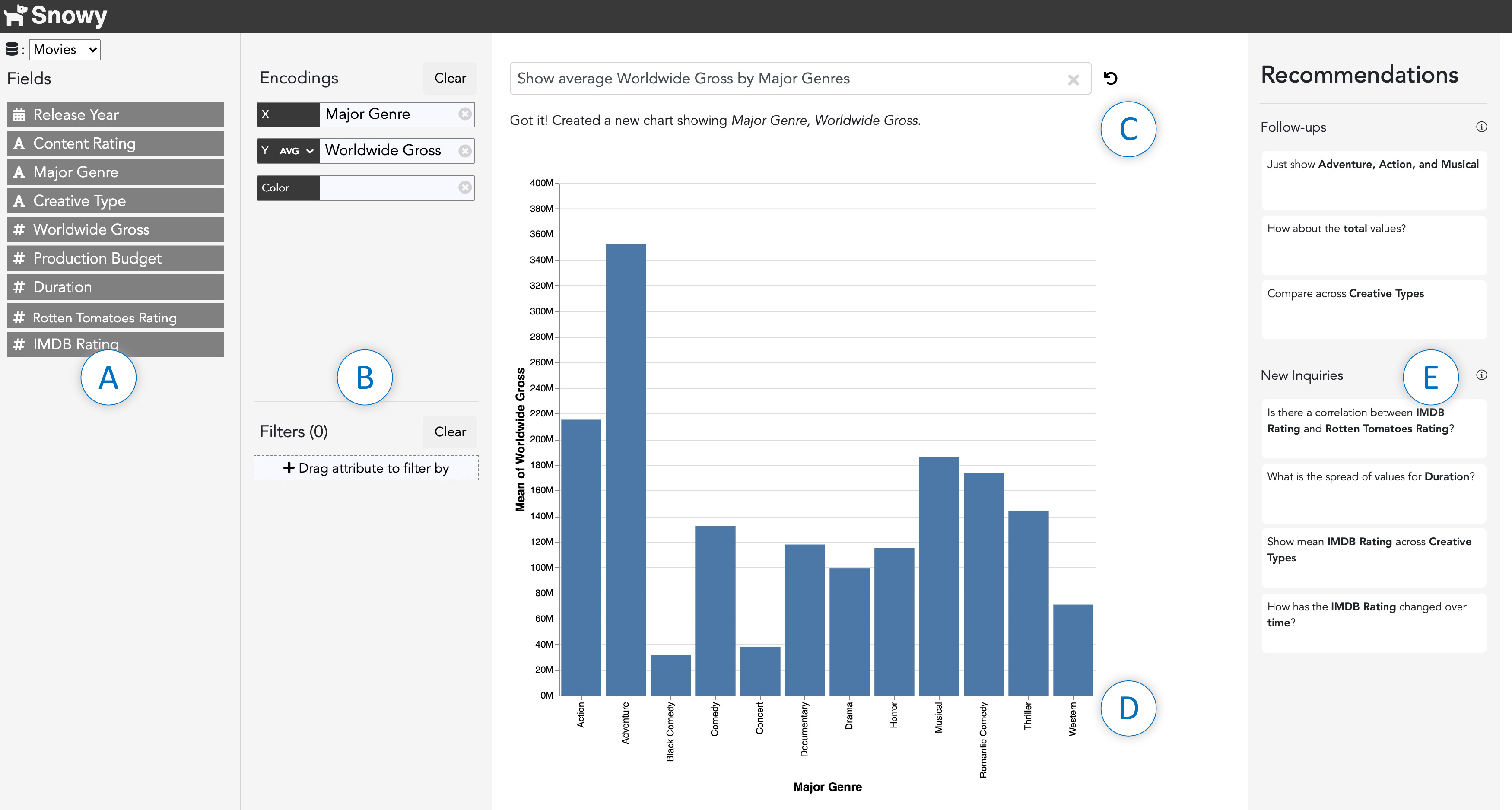}
    \caption{\snowy's user interface while exploring a movies dataset.
    (A) Attribute panel, (B) manual view specification and filtering panel, (C) NL input and feedback, (D) visualization canvas, and (E) recommendations panel.
    Here, given the input utterance ``\textit{Show average Worldwide Gross by Major Genres}'', \snowy~creates a bar chart as a response and updates the encodings panel to reflect the current view.
    Simultaneously, several utterance recommendations are also presented to suggest modifying the current chart (Follow-ups), as well as for exploring other aspects of the dataset (New Inquiries).
    Users can click a recommendation to submit it as their input utterance, or right-click to copy the recommendation's text into the input box and edit it further.}
    \label{fig:interface}
    \Description{Figure 2 shows Snowy's user interface.}
\end{figure*}

\section{Snowy}

Incorporating the aforementioned design considerations, we developed \snowy~as a prototype system to investigate the idea of leveraging utterance recommendations for guiding visual analysis, while implicitly helping users discover and learn the system's NL capabilities.
In this section, we first briefly describe \snowy's interface and walk through a usage scenario that exemplifies how utterance recommendations can aid conversational visual analysis.
We subsequently detail \snowy's design and implementation, discussing how it leverages a combination of features from the underlying data along with the users' interaction context to present in-situ recommendations.

\subsection{Interface and Usage Scenario}
\label{sec:interface-and-scenario}

\snowy~(Figure~\ref{fig:interface}) is a mixed-initiative visualization system that supports NL input as well as direct manipulation interactions through a graphical user interface (GUI).
Let us now consider an exemplary usage scenario to understand how the different components of \snowy's interface collectively support visual analysis (this scenario is also illustrated in the supplementary video).

\begin{figure}[t!]
    \centering
    \includegraphics[width=.9\linewidth,keepaspectratio]{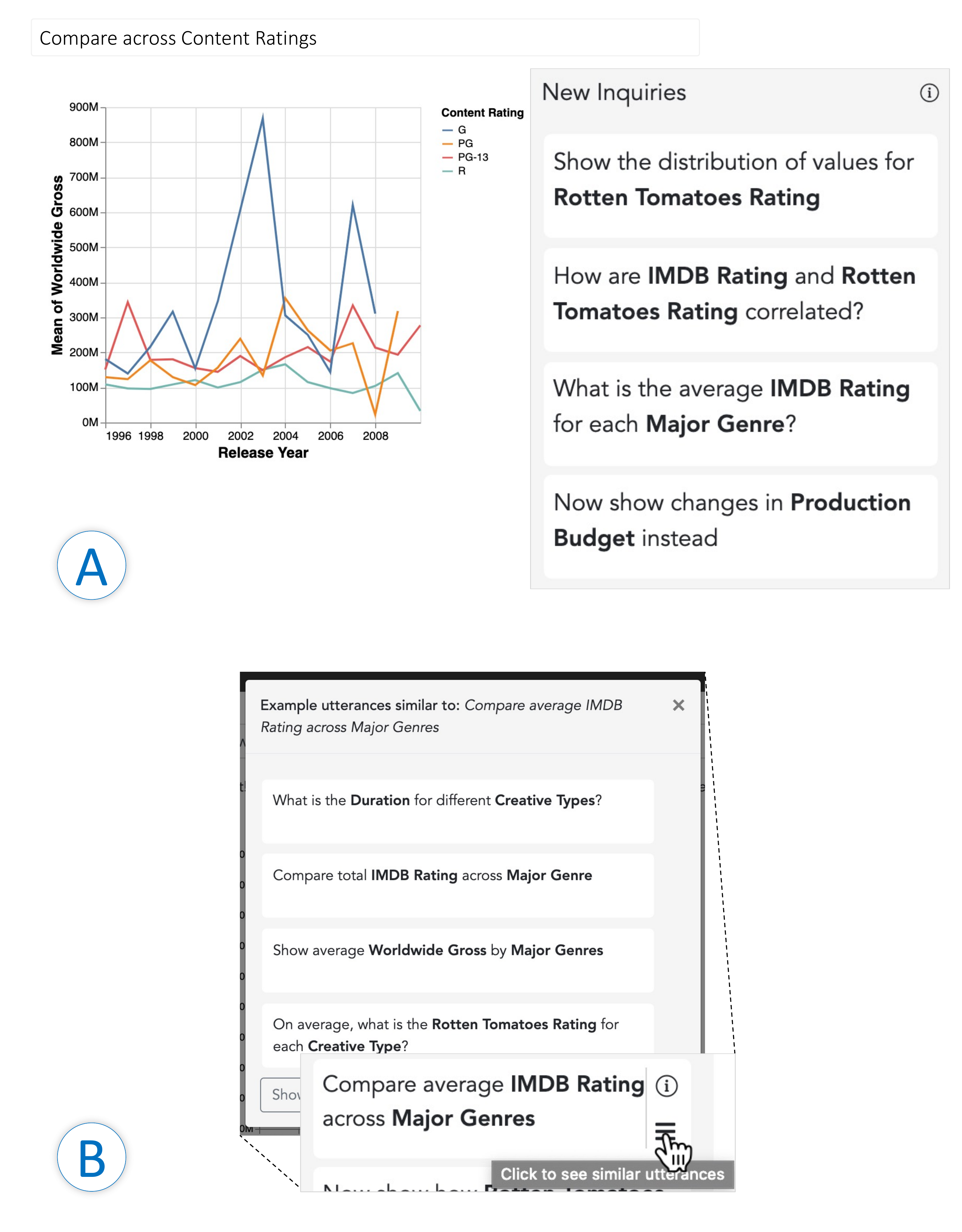}
    \caption{Tintin's exploration of the movies dataset. (A) The initial line chart in Figure~\ref{fig:teaser}B is updated to group lines by \textit{Content Rating} and new recommendations are shown. Notice that the phrasing of the last recommendation reflects the system's support for pragmatics and is phrased in linguistic continuation with the previous utterances. (B) Alternative utterance examples are suggested based on an initial recommendation.}
    \label{fig:scenario-1}
    \Description{Figure 3 shows two screenshots from the illustrative usage scenario of exploring a movies dataset. Figure 3A displays recommendations highlighting how pragmatic markers such as ``instead" can be used to replace active attributes in the visualization. Figure 3B shows recommendations similar to ``Compare average IMDB Rating across Major Genres".}
\end{figure}


Tintin, an analyst at a movie production house is examining a dataset of previously released movies to identify the types of movies his company should invest in next.
The dataset\footnote{Provided as supplementary material.} contains $700$ movies with nine attributes for each movie including its {\small{\faCalendar}}~\textit{Release Year}, {\small{\faFont}}~\textit{Major Genre}, {\small{\faHashtag}}~\textit{Worldwide Gross}, and {\small{\faHashtag}}~\textit{IMDB Rating}, among others (shown in Figure~\ref{fig:interface}A).
For consistency, we use this movies dataset for examples throughout this paper.

As Tintin loads the dataset, \snowy~scans the underlying data to identify potentially interesting attribute combinations to explore and presents a list of utterance recommendations that Tintin can use to start his exploration (Figure~\ref{fig:teaser}A) (\textbf{DG4}). Tintin finds the recommendation \reco{What is the trend of \textbf{Worldwide Gross} over the \textbf{years}?} to be relevant for observing trends and clicks the recommendation to select it as his input utterance.
In response, \snowy~renders a line chart and then updates its recommendations to suggest new inquiries, while also including some new recommendations of follow-up utterances to modify the active chart (Figure~\ref{fig:teaser}B) \textbf{(DG2)}.
Looking through the follow-up recommendations, Tintin is intrigued by the idea of continuing to observe the trend of budget but across different \textit{Content Ratings}, and selects the recommended utterance \reco{Compare across \textbf{Content Ratings}}.
Viewing the updated multi-series line chart, Tintin observes that while all movie types except R-rated movies have a higher gross over time, PG-13 movies have had the most stable increase (Figure~\ref{fig:scenario-1}A).
Tintin again looks at the new inquiries suggested by \snowy~and selects the recommendation \reco{Now show changes in \textbf{Production Budget} instead} to contrast trends across \textit{Worldwide Gross} and \textit{Production Budget}.
In response, \snowy~updates the line chart, swapping the \textit{Worldwide Gross} attribute on the Y-axis with the \textit{Production Budget}.
Through the updated chart, Tintin notices that movies follow a similar trend for the budget and correspondingly makes a note that his company should consider PG-13 movies more closely.

\begin{figure}[b!]
    \includegraphics[width=.9\linewidth,keepaspectratio]{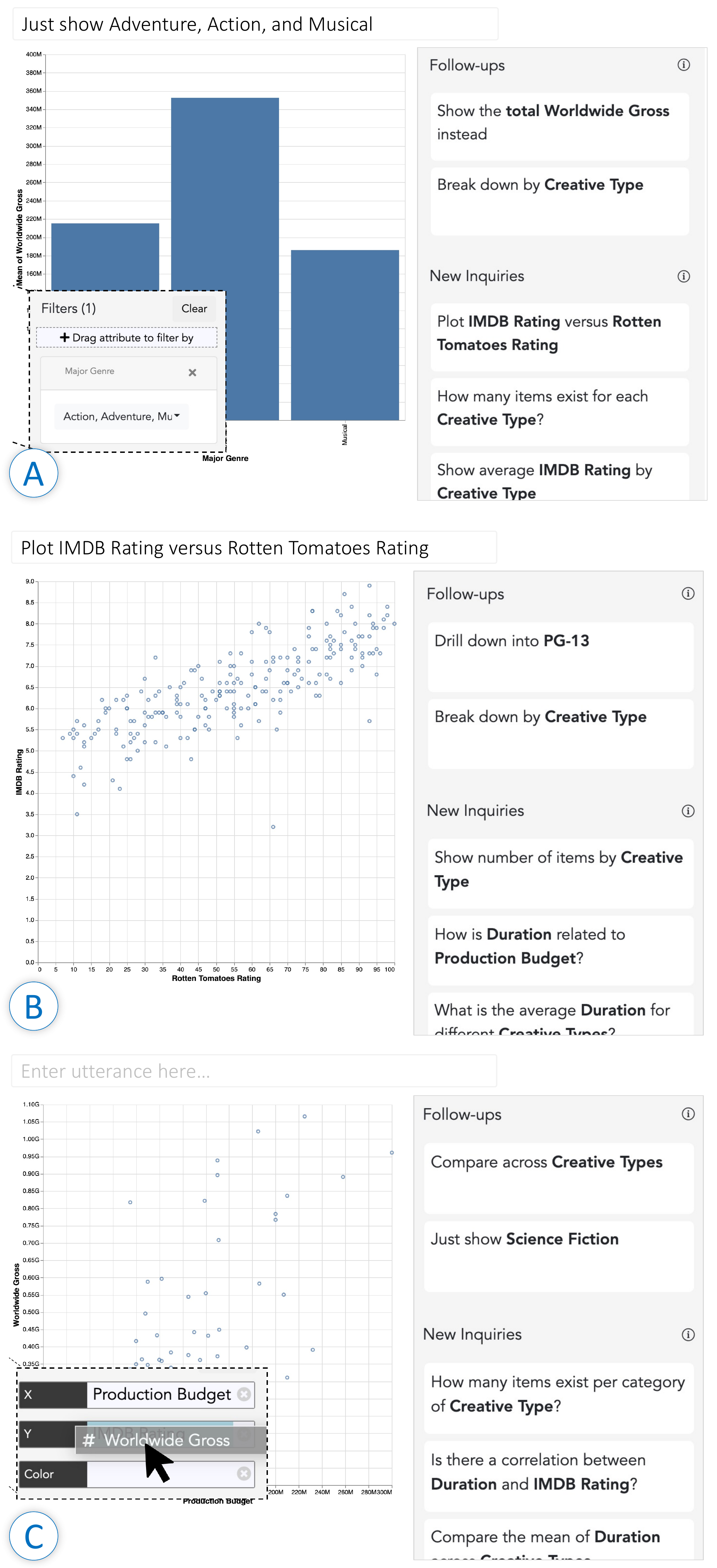}
    \caption{Tintin's exploration of the movies dataset (cont'd). (A) \textit{Major Genre} is applied as a filter, along with a dropdown to optionally refine the filter.
    (B) A new scatterplot is created while preserving the filters from the bar chart.
    (C) The input box is cleared as encodings are updated through the manual view specification. New recommendations based on the updated chart are shown.}
    \label{fig:scenario-2}
    \Description{Figure 4 shows three screenshots from the illustrative usage scenario of exploring a movies dataset. Figure4A shows how a Major Genre filter applied through natural language utterances is also added to the graphical user interface for further refinement. Figure 4B shows how a new scatterplot is created while preserving filters from the previous bar chart. Figure 4C displays that the input query box is cleared when the visualization is updated through the manual view specification panel.}
\end{figure}

Wondering which attributes to explore next, Tintin skims the recommendations panel and pauses when he sees the utterance \reco{Compare average \textbf{IMDB Rating} across \textbf{Major Genres}}.
He likes the idea of comparing values across groups of genres, but does not want to use the \textit{IMDB Rating}s.
To see alternative recommendations, he hovers on the recommendation and then clicks the~{\small{\faBars}}~icon to see similar recommendations (Figure~\ref{fig:scenario-1}B).
From the suggested list, he selects \reco{Show average \textbf{Worldwide Gross} by \textbf{Major Genre}} as his input utterance. Inspecting the resulting bar chart, Tintin observes that \textit{Action} and \textit{Adventure} have notably higher values. In parallel, \snowy~also interprets the bar chart and recommends a series of follow-up recommendations, including one to drill down into the three highest grossing genres (\reco{Just show \textbf{Adventure, Action, and Musical}}). As this recommendation matches his observation, Tintin selects this follow-up utterance to refine the scope of his exploration. The system adds a \textit{Major Genre} filter, allowing Tintin to refine it further through the GUI if needed (Figure~\ref{fig:scenario-2}A).

Considering the active bar chart, Tintin's prior interactions, and the available data attributes, \snowy~now recommends two follow-up utterances to include \textit{Creative Type} as an additional attribute for comparison, or change the aggregation level of the current chart from \textit{average} to \textit{sum}.
\snowy~also infers that Tintin has performed analytic tasks like looking for trends over time and compared values for quantitative attributes across groups of categorical values.
Thus, to broaden his analytic coverage, \snowy~promotes utterances pertaining to other analytic tasks like observing correlations between attributes and seeing the distribution items in the dataset in its new inquiry recommendations (Figure~\ref{fig:scenario-2}A) (\textbf{DG1,DG3}).
Furthermore, to promote data attribute coverage in his exploration, \snowy~populates these recommendations with attributes that Tintin has not previously considered (e.g., \textit{Creative Type}, \textit{IMDB Rating}, \textit{Duration}) (\textbf{DG1}).

Seeing the utterance recommendation \reco{Plot \textbf{IMDB Rating} verus}\\ \reco{\textbf{Rotten Tomatoes Rating}} at the top of the new inquiry recommendations, Tintin realizes that he has not considered movie ratings as part of his analysis so far and selects it as his input, resulting in \snowy~rendering a scatterplot comparing ratings across the two platforms (Figure~\ref{fig:scenario-2}B).
Tintin subsequently selects the follow-up recommendation \reco{Drill down into \textbf{PG-13}} to further filter down to \textit{PG-13} movies (because he had earlier noticed that these movies tend to have a stable return on investment over time).
Inspecting the scatterplot, Tintin confirms that the ratings across the IMDB and Rotten Tomatoes platforms are fairly correlated and decides to switch his focus back to the return on investment on movies. 
To do so, he now manually updates the scatterplot by dragging the \textit{Production Budget} and \textit{Worldwide Gross} attribute pills to the X- and Y-encoding shelves, respectively.
In response, \snowy~clears the text input box since the visualization was manually specified and updates its utterance recommendations based on the new chart (Figure~\ref{fig:scenario-2}C).
Tintin selects the follow-up recommendation \reco{Compare across \textbf{Creative Types}} since that is an attribute he had not previously considered, leading to the system coloring points by \textit{Creative Type}.

To inspect movies with a high gross and return on investment, Tintin draws a selection box on the chart around movies that gross over $\sim$\$500 million while having a budget of $\sim$\$200 million or below.
Through the active selection of 13 movies and their mark colors, Tintin identifies that \textit{Science Fiction} (five movies), \textit{Super Hero} (four movies) have the highest chance of success, with \textit{Contemporary Fiction} and \textit{Fantasy} (two movies each) also being creative types to consider.
Based on Tintin's active selection, \snowy~now suggests follow-up utterances for computations that can be performed on the selected items (Figure~\ref{fig:teaser}C) (\textbf{DG2}).
Tintin selects the recommendation \reco{What are the \textbf{average} values?} and notes that PG-13 movies having the specified genres and creative types gross, on average, \$709 million on an average investment of \$132 million.
He clears the view and continues exploring other aspects of the data to report any additional findings to his company's investment team.
\subsection{System Overview}

\begin{figure}[b!]
    \includegraphics[width=\linewidth,keepaspectratio]{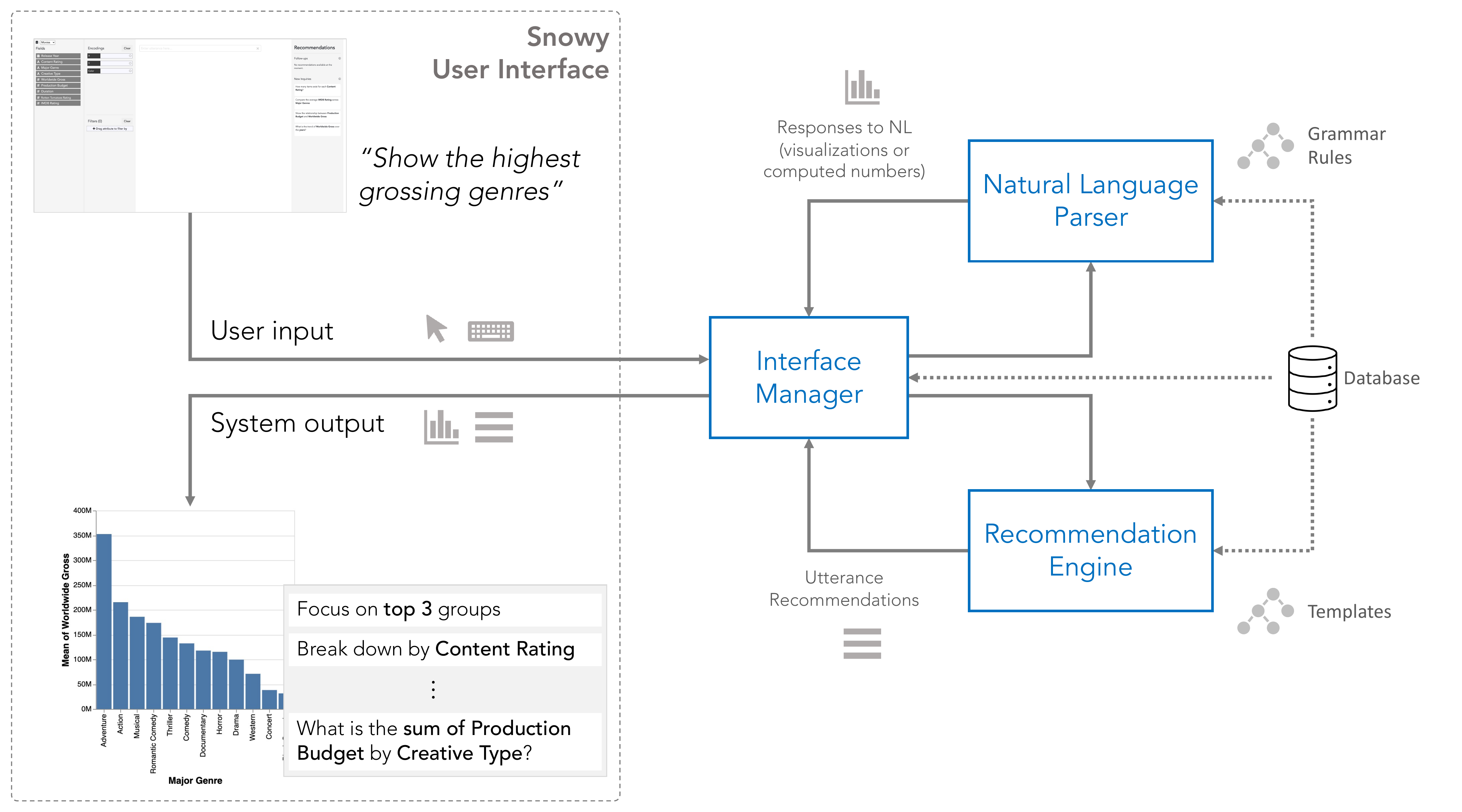}
    \caption{System architecture overview.}
    \label{fig:architecture}
    \Description{Figure 5 shows Snowy's system architecture, highlighting connections between the main components---namely, the interface manager, the natural language parser, and the recommendation engine.}
\end{figure}

\snowy~is implemented as a web-based application and is developed using Python, HTML/CSS, and JavaScript. \edited{The system accepts any tabular CSV dataset as input.} Visualizations in the tool are specified and created using Vega-Lite~\cite{satyanarayan2016vega}.
\snowy~currently supports three encoding channels (\texttt{x}, \texttt{y}, \texttt{color}) and three mark types (\texttt{bar}, \texttt{line}, \texttt{point}).
These marks and encodings collectively allow users to specify and interact with a canonical set of visualizations including bar charts, line charts, and scatterplots \edited{that cover the range of analytic intents currently supported in \snowy.}
During manual view specification, \snowy~selects the default visualization using a simplified version of the Show Me system~\cite{mackinlay2007show}, employing similar rules to determine mark types based on the mappings between the visual encodings and attribute data types (e.g., showing a scatterplot if two quantitative attributes are mapped to the \texttt{xy}-channels and showing a line chart if a temporal attribute is visualized on the \texttt{x}-axis with a quantitative attribute on the \texttt{y}-axis).

Figure~\ref{fig:architecture} presents a high-level depiction of the system's architecture.
There are three main components---namely, the \textit{Interface Manager}, the \textit{Natural Language Parser}, and the \textit{Utterance Recommendation Engine}.
In the following sections, we describe these individual components and highlight how they collectively support the features in \snowy.
\subsection{Interface Manager}
\label{sec:interface-manager}
\snowy~is designed to be a context-sensitive recommendation interface that employs an action-reaction design~\cite{context-modeling:2007}, where the user interaction causes the interface to react and update based on the actions that the user makes.
User actions that drive recommendations in \snowy~include manual view specification or filtering through drag-and-drop, typing NL utterances, selecting recommended utterances, and selecting marks in the active visualization.

As users interact with the tool, \snowy~tracks their actions and maintains a \textit{context state object} that drives the system's utterance recommendations.

\begin{figure*}[t!]
    \centering
    \includegraphics[width=\textwidth]{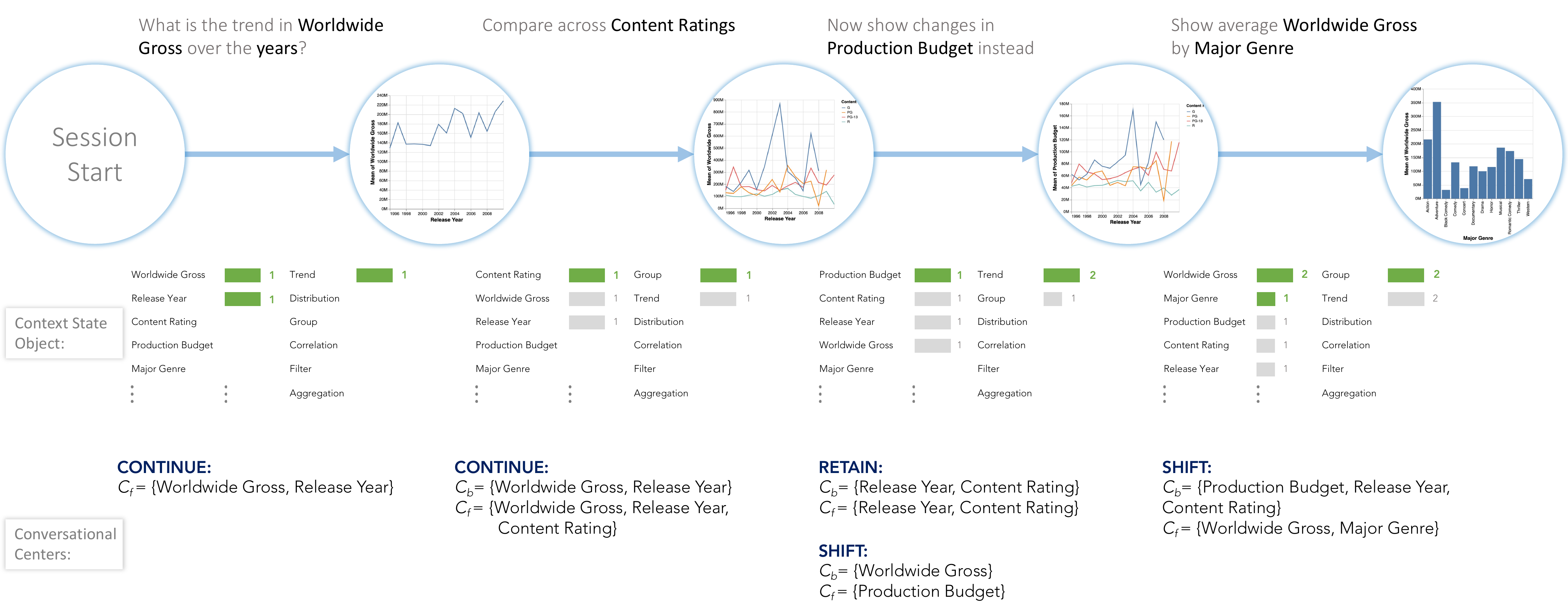}
    \caption{Progression of the context state and conversational centers (with \textcolor{transitionBlue}{TRANSITION types \textbf{in blue}}) through the initial stages of the usage scenario described in Section~\ref{sec:interface-and-scenario}.
    For the context state object, attributes and intent interaction scores are shown below each transition; score increments are highlighted \textcolor{incrementGreen}{\textbf{in green}} \edited{and are also represented by the bar lengths}.
    The final state (bar chart of \textit{Worldwide Gross} by \textit{Major Genre}) in this figure corresponds to the interface state shown in Figure~\ref{fig:interface}.}
    \label{fig:context-state-changes}
    \Description{Figure 6 demonstrates a sub-sequence from the illustrative usage scenario showing the changes in attribute and intent scores in Snowy's context state object along with how the conversation centers are maintained.}
\end{figure*}

\subsubsection{Modeling the context state}
The context state object is modeled as a combination of 1) the active chart and mark selections, 2) the active utterance, and 3) the user's data exploration history.
The first two are directly captured from the visualization canvas (Figure~\ref{fig:interface}D) and the input box (Figure~\ref{fig:interface}C) respectively.
The data exploration history is stored as interaction scores corresponding to attributes, values, and the various analytic intents supported in \snowy~(i.e., \intent{distribution}, \intent{group},  \intent{correlation}, \intent{trend}, \intent{filter}, and \intent{aggregation}).
While this is not a definitive general model for capturing and understanding user interest, it helps track the user's coverage in the context of data exploration, enabling the system to promote depth- and/or breadth-oriented exploration through its recommendations (\textbf{DG1}).

For attributes and values, the interaction scores are computed by tracking the number of times data attributes are mapped to visual encodings or data values applied as filters.
For intents, the scores are incremented in three scenarios.
First, when utterance recommendations are selected, the score for the recommendation's corresponding intent is incremented (e.g.~selecting the recommendation \reco{Drill down into \textbf{PG-13}} in Figure~\ref{fig:scenario-2}B increments the score for the \intent{filter} intent).
Second, when NL utterances are manually entered, \snowy~infers the underlying intent from the input utterance and increments the score based on its confidence in the interpretation.
For example, an input utterance like ``\textit{Show the correlation between IMDB Ratings and Production Budget}" leads to a higher increment for the \intent{correlation} intent compared to the utterance ``\textit{Show IMDB Ratings against Production Budget}" since there is an explicit reference to the intent in the first utterance versus an implicit one in the second.
And third, specifically for the \intent{filter} intent, the score is also incremented if new filters are manually applied through the GUI.

Figure~\ref{fig:context-state-changes} shows instances of the context state object during the aforementioned usage scenario.
Between the first and third states in Figure~\ref{fig:context-state-changes}, for example, two recommendations were selected (\reco{What is the trend of \textbf{Worldwide Gross} over the \textbf{years}?}, \reco{Compare ac-} \reco{ross \textbf{Content Ratings}}).
This leads to the system \edited{incrementing} the interaction scores for the attributes \textit{Worldwide Gross}, \textit{Release Year}, and \textit{Content Rating}, as well as the intents \intent{trend} and \intent{group} (based on the first and second utterance, respectively) \edited{from 0 to 1}.
\subsection{Natural Language Parser}
\label{sec:nl-parser}

The parser to interpret NL utterances is implemented using the open-source Natural Language for Data Visualization (NL4DV) toolkit~\cite{narechania2020nl4dv}.
The toolkit takes as input a dataset and an utterance corresponding to that dataset, returning a JSON object that includes 1) the attributes and intents referred to in the utterance and 2) an ordered list of Vega-Lite~\cite{satyanarayan2016vega} specifications that can be presented in response to the input utterance.
We chose to build upon NL4DV instead of implementing a fully-custom module since language understanding was not our primary research focus.
However, while NL4DV provides basic interpretation capabilities and the ability to specify a visualization through NL, the toolkit does not support conversational interaction through follow-up and deictic utterances, or utterances for statistical computations (e.g., computing differences or correlation coefficients).

We extend NL4DV's default grammar with a set of production rules augmented with both syntactic and semantic predicates based on analytical expressions that correspond to different intents commonly found in mainstream systems like Tableau's VizQL~\cite{polaris,tableau} and Ask Data~\cite{askdata}.
We add support for follow-up utterances by adding a `follow-up' grammar rule that is triggered by pragmatic markers.
Pragmatic markers are linguistic cues that convey the intentionality of a follow-up utterance in reference to the current context~\cite{fraser:1990}.
These pragmatic markers are often adverbs (e.g., `also', `how about`) or referentials (e.g., `this', `that') that signal the user’s potential communicative intention of following up from a previous conversation state.

Listing~\ref{listing:grammar} shows a subset of the underlying grammar with its production rules specified for the various analytical intents and follow-up utterances. In the grammar, fields without an aggregation are called \emph{dimensions} (typically,~{\small{\faFont}}~categorical and~{\small{\faCalendar}}~temporal attributes) , while \emph{measures} (typically,~{\small{\faHashtag}}~quantitative attributes) are fields that are aggregated within groups defined by the set of all dimensions. For brevity, we only \edited{show a subset of the production rules and} excluded synonyms, date, and place terminals from the grammar in Listing~\ref{listing:grammar}.


\begin{figure}
\centering
\begin{minted}
[
baselinestretch=1.3,
frame=lines,
fontsize=\footnotesize,
% xleftmargin=15pt,
breaklines,
style=manni,
escapeinside=@@,
mathescape=true
]{javascript}

<utterance> @$\rightarrow$@ <groupCmd> | <distributionCmd> |  <correlateCmd> | <trendCmd> | <filterCmd> | <extremaCmd> | <followUpCmd>;
<groupCmd> @$\rightarrow$@ <dimension> ('group by' | 'by')? <aggMeasure>;
<aggMeasure> @$\rightarrow$@ aggTerms <measure>;
<aggTerms> @$\rightarrow$@ ('average' | 'median' | 'mean' | 'min' | 'max');
<distributionCmd> @$\rightarrow$@ ('distribution' | 'bin') <measure>;
<correlateCmd> @$\rightarrow$@ <measure> ('correlation' | 'scatterplot' | 'relationship')? <measure>;
<trendCmd>  @$\rightarrow$@ ('trend' | 'over time')? <dimension> <dateAttribute>;
<filterCmd>  @$\rightarrow$@ <locationFilter> | <temporalFilter> | <valueFilter>;
<locationFilter> @$\rightarrow$@ filterPlaceCmd  <location>;
<temporalFilter> @$\rightarrow$@ ('in' | 'before' | 'after') <timeDataValue>| <timeDateRange>;
<valueFilter> @$\rightarrow$@  (<lessThan> | <equalTo> | <greaterThan> | <lessEqual> | <greaterEqual>) <measure>;
<extremaCmd>  @$\rightarrow$@ ('highest' | 'largest' | 'smallest' | 'lowest') <measure>;
<calculationCmd>  @$\rightarrow$@ ('correlation' | 'difference' | 'total' | 'lowest' | aggTerms) <measure>;
<followupCmd>  @$\rightarrow$@ (<pragmaticMarker> | <referential>) <utterance>;
<pragmaticMarker> @$\rightarrow$@ (('what'| 'how') 'about') | 'also' | 'just' | 'only')?;
<referential>  @$\rightarrow$@ ('this' | 'that');}
\end{minted}
\vspace{-1em}
\captionof{listing}{Subset of grammar production rules.}
\vspace{-1em}
\label{listing:grammar}
\end{figure}

\subsubsection{Supporting a conversational model}
To support follow-up utterances, we apply principles of pragmatics by modeling the interactions and recommendation behaviors as a conversation.
Specifically, we incorporate the conversational interaction model for visual analysis proposed by Hoque et al.~\cite{hoque2017applying}.
This model builds upon a conversational centering approach~\cite{Grosz:1986}, where utterances are divided into constituent conversational segments, embedding relationships that may hold between segments.
A center $C$ refers to those entities serving to link that utterance to other utterances in the conversation. For a total of $m$ utterances in a conversation, each utterance $U_n$ ($1 \leq n < m$) in the conversation $converse$ is assigned a set of forward-looking centers, $C_f(U_n, converse)$ referring to the current focus of the conversation; each utterance other than the initial utterance, is assigned a set of backward-looking centers, $C_b(U_n, converse)$, referring to the previous state of the conversation.
The forward and backward-looking centers consist of data attributes and values, visual properties, and analytical intent.
Transitions from the backward-looking center to the forward-looking center are realized through three types of transition states: \\
\textbf{Continue}: Transition that continues the context from the backward-looking center to the forward-looking one, while potentially adding new entities. \\
\textbf{Retain}: Transition retains the context from the backward-looking center in the forward-looking one \textit{without} adding additional entities to the forward-looking one.\\
\textbf{Shift}: Transition shifts or changes context from the previous one.
\newline

\noindent{}Based on this transition model, given an utterance $U_n$, the parser responds by executing a series of analytical functions derived from the forward-looking centers $C_f(U_n, converse)$. Figure~\ref{fig:context-state-changes} illustrates these different types of transitions during the conversation between the movie production house analyst, Tintin and our system, \snowy.
In the example, the first utterance asking about gross over years sets the conversational center to the \textit{Worldwide Gross} and \textit{Release Year} attributes.
The subsequent utterance asking for a comparison across groups adds an attribute, \textit{Content Rating} to the current center, resulting in a \textit{continue} transition.
The third utterance includes a pragmatic marker (`instead') and requests for a change in the attribute shown on the active line chart and indicates interest in the \intent{trend} intent.
Referring to the existing center, the system swaps out the \textit{Worldwide Gross} measure for the \textit{Production Budget}, performing a \textit{shift} transition, while \textit{retaining} the other two attributes.
Finally, the next utterance ``\textit{Show average Worldwide Gross by Major Genre}'' switches to a new set of attributes and intent (\intent{group}), and thus \textit{shifts} the center to the \textit{Worldwide Gross} and \textit{Major Genre} attributes, as the system creates a new bar chart.

\subsubsection{\edited{Error Handling}}
\edited{Similar to other visualization NLIs (e.g.,~\cite{sun2010articulate,datatone,setlur2016eviza}), \snowy~also encounters errors when parsing NL input: (1) utterances with ambiguous references (e.g., `rating' can map to multiple attributes, \textit{Content Rating} and \textit{IMDB Rating}), (2) underspecified utterances (e.g.~``\textit{imdb ratings by genre}'' does not specify a chart type or if \textit{IMDB Ratings} should be aggregated), and (3) utterances for unsupported operations such as formatting (e.g.~``\textit{Change blue bars to red}'').}

\edited{Although the system does not update the visualization for utterances requesting unsupported operations, ambiguous and underspecified utterances are handled internally by NL4DV~\cite{narechania2020nl4dv}, which selects reasonable defaults.
However, \snowy’s direct manipulation interface enables users to override these defaults (e.g.~dragging an attribute to manually set a binding to override the system default in the case of an utterance with an ambiguous attribute reference).
Additionally, the feedback below the text input box also displays potential errors and provides an {undo ({\small{\faUndo}}) option} to revert the last utterance (Figure~\ref{fig:interface}).
Note, however, that the ambiguities and errors do not occur when recommended utterances are selected since \snowy~is fully aware of the features driving the utterances and only recommends unambiguous phrasings.}
\subsection{Utterance Recommendation Engine}
\label{sec:reco-engine}

Central to \snowy~is its recommendation engine that generates contextual utterance suggestions.
The system generates the recommendations by considering a combination of patterns in the underling data (e.g., strong correlations, variations in values over a temporal attribute), a user's session history (e.g., attributes considered, filters applied), and any active interactions with the interface. User interactions that trigger recommendations include issuing an NL utterance, updating the visualization through the manual view specification and filtering panel, as well as directly selecting marks on the active visualization, a behavior commonly referred to as \emph{deictic referencing}~\cite{CLARK1983245}.

Figure~\ref{fig:recommendation-example} provides an overview of \snowy's recommendation engine. The engine takes the context state object as input from the interface manager and returns an ordered list of utterance objects in response (Figure~\ref{fig:recommendation-example}-top).
Utterance objects contain the recommendation text along with other meta-information including the utterance type (`Follow-up' versus `New Inquiry') and the associated intent (e.g., \intent{filter}, \intent{group}).
Given a context state, \snowy~performs three steps to generate utterance recommendations: 1) \textit{filtering and ranking}, 2) \textit{parameterization}, and 3) \textit{linguistic realization}.
Figure~\ref{fig:recommendation-example} provides a summary of these steps using a system state from the aforementioned scenario (Figure~\ref{fig:interface}).
We use Figure~\ref{fig:recommendation-example} as a running example for the remainder of this section.

\subsubsection{Filtering and Ranking}

\begin{figure*}[t!]
    \centering
    \includegraphics[width=.95\textwidth]{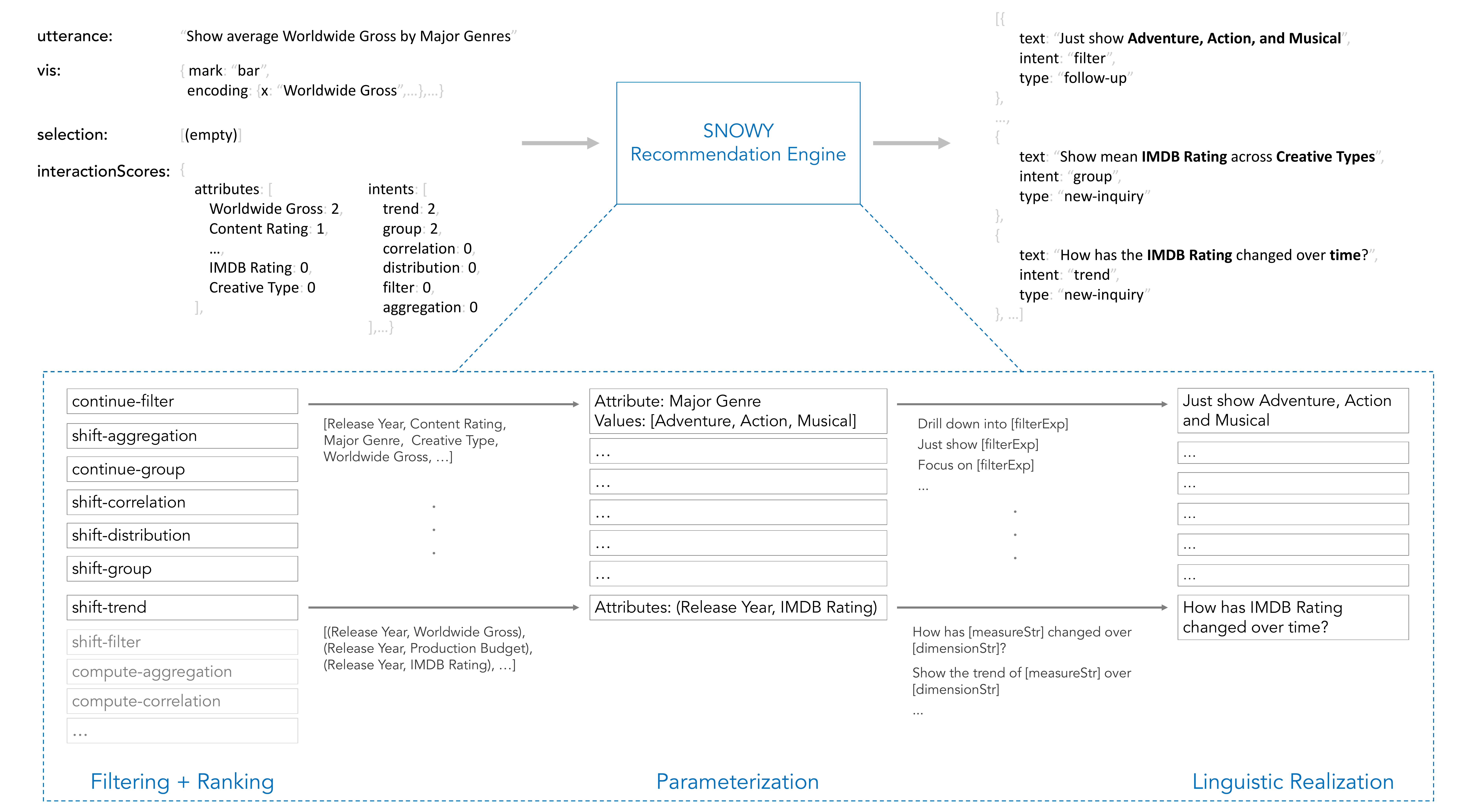}
    \caption{An overview of \snowy's recommendation engine using the system state shown in Figure~\ref{fig:interface} as an example.
    (Top) Given a context state object as input, the engine generates a list of utterance recommendations to suggest in \snowy's UI.
    (Bottom) Steps executed to generate the recommendations: (1) filtering + ranking, (2) parameterization, and (3) linguistic realization of recommendations.}
    \label{fig:recommendation-example}
    \Description{Figure 7 gives an overview of Snowy's recommendation engine.}
\end{figure*}

\snowy~uses the context state to determine which types of utterances it should recommend at any point in time. Specifically, if one or more marks are selected in the active visualization, \snowy~prioritizes deictic utterances and checks if it can recommend utterances based on the current chart type and selection. For instance, in Figure~\ref{fig:teaser}C, upon detecting multiple selected marks in a scatterplot, \snowy~recommends \reco{What are the \textbf{average} va-} \reco{lues?} and \reco{What is the \textbf{correlation} between these points?} as follow-up utterances since computing averages and correlations are common analytic tasks for scatterplots~\cite{sarikaya2017scatterplots}.

Alternatively, if there are no mark selections but there is an active chart in play, \snowy~checks for follow-up utterances to recommend.
Follow-up recommendations typically promote \textit{continue} or \textit{shift} transitions through utterances that correspond to: 1) adding one or more new attributes to the view if there are unused encoding channels (e.g.~\reco{Compare across \textbf{Creative Types}} in Figure~\ref{fig:interface} since the \textit{color} channel is available), 2) changing the active chart's aggregation level (e.g.~\reco{How about the \textbf{total} values?} in Figure~\ref{fig:interface} as the bar chart shows the \textit{mean}), and 3) adding filters (e.g.~\reco{Just show \textbf{Adventure, Action, and Musical}} in Figure~\ref{fig:interface} for the active \textit{Major Genre} attribute).

Besides follow-up utterances for an active chart or selection, \snowy~also generates new inquiry recommendations to suggest alternative analytic paths for users to consider (\textbf{DG2}). These recommendations focus on \textit{shift} transitions in the conversational interaction model and map to one or more analytic intents (\textbf{DG3}).
Examples of new inquiry recommendations in Figure~\ref{fig:interface} include \reco{What is the spread of values for \textbf{Duration}?} and \reco{Show mean \textbf{IMDB Ra-}} \reco{\textbf{ting} across \textbf{Creative Types}}, that suggest shifting focus to the underexplored \textit{Duration}, \textit{IMDB Rating}, and \textit{Creative Type} attributes while considering \intent{distribution} and \intent{group} as the underlying analytic intents, respectively.

After shortlisting recommendations types, \snowy~orders them such that recommendations mapping to the least covered intent-types (determined using the context state object) precede other recommendations.
This ordering helps promote breadth in analytic querying and also tries to make users aware of potentially underexplored system capabilities given their prior interactions (\textbf{DG1}).
In the context of Figure~\ref{fig:recommendation-example}, this ordering logic results in \textit{continue-filter} and \textit{shift-aggregation} being shown before \textit{continue-group} since the user had previously issued a \textit{continue-group} utterance (\reco{Compare across \textbf{Content Ratings}}, Figure~\ref{fig:context-state-changes}).

\subsubsection{Parameterization}
\label{sec:parameter-assignment}

\begin{table*}[t!]
\centering
\resizebox{\textwidth}{!}{%
\begin{tabular}{@{}llll@{}}
\toprule
\begin{tabular}[c]{@{}l@{}} \textbf{Intent} \\ \\ \\ \end{tabular} &
  \begin{tabular}[c]{@{}l@{}} \textbf{Parameters} \\ \\ \\ \end{tabular} &
  \begin{tabular}[c]{@{}l@{}}\textbf{Parameter Selection Functions}\\ {\small{(used in combination with the interaction}}\\{\small{scores from the context state object)}}\end{tabular} &
  \begin{tabular}[c]{@{}l@{}} \textbf{Parameter Selection Logic} \\ \\ \\ \end{tabular} \\ \midrule
Correlate &
  (measure, measure) &
  \begin{tabular}[c]{@{}l@{}}Pearson's correlation\\ coefficient ($r$)\end{tabular} &
  \begin{tabular}[c]{@{}l@{}}Attribute combinations that have higher $|r|$ are prioritized over combinations with\\ lower $|r|$.\end{tabular} \\
 &
   &
   &
   \\
Group &
  (dimension, measure) &
  Standard deviation ($\sigma$) &
  \begin{tabular}[c]{@{}l@{}}Attribute combinations with higher $\sigma$ are prioritized over combinations with lower\\ $\sigma$. By default, $\sigma$ is calculated using the mean values for a \texttt{measure} over each\\ group/category in a \texttt{dimension}.\end{tabular} \\
 &
   &
   &
   \\
Trend &
  (dimension, measure) &
  Standard deviation ($\sigma$) &
  \begin{tabular}[c]{@{}l@{}}Attribute combinations with higher $\sigma$ are prioritized over combinations with lower\\ $\sigma$. By default, $\sigma$ is calculated using the mean values for a \texttt{measure} over each\\ timestamp in a \texttt{dimension}.\end{tabular} \\
 &
   &
   &
   \\
Distribution &
  (dimension, measure) &
  Standard deviation ($\sigma$) &
  \begin{tabular}[c]{@{}l@{}}Attribute combinations with higher $\sigma$ are prioritized over combinations with lower\\ $\sigma$. $\sigma$ is calculated using the number of items in each group/timestamp for a \\ \texttt{dimension} and bins of values in a \texttt{measure}.\end{tabular} \\
 &
   &
   &
   \\
Filter &
  \begin{tabular}[c]{@{}l@{}}(measure, value range),\\ (dimension, categories),\\ (dimension, time range)\end{tabular} &
  \begin{tabular}[c]{@{}l@{}}Top N, \\ Pearson's correlation\\ coefficient (r),\\ Standard deviation ($\sigma$)\end{tabular} &
  \begin{tabular}[c]{@{}l@{}}For categorical dimensions, select groups/categories with highest values for the\\ item count or measure in the active visualization (for bar charts), $|r|$ (for scatterplots),\\and $\sigma$ (for line charts).\\ For measures and temporal dimensions, first, identify groups using quartile ranges.\\Then, compute $|r|$ or $\sigma$ for the measures in the active visualization. Numeric and\\ temporal filters do not get suggested for bar charts since bar charts show aggregated\\ information and the effects of these filters may not be directly evident perceptually.\end{tabular} \\ \bottomrule
\end{tabular}%
}
\vspace{.75em}
\caption{Parameters required to populate different types of utterance recommendations along with the underlying parameter selection logic. Besides the five listed categories, \snowy~also generates \intent{aggregation} change recommendations. However, these have a fixed set of parameter values (either \textit{mean} or \textit{sum} in the current prototype).}
\label{tab:parameter-selection}
\Description{Table 1 lists the supported intent types (correlate, group, trend, distribution, filter), their parameters (attributes and/or values), and the logic used to select the parameters.}
\end{table*}
As a next step, the system needs to parameterize these shortlisted recommendations with appropriate data and analytic features including attributes (e.g., \textit{Content Rating}, \textit{Worldwide Gross}), values (e.g., \textit{PG-13}, \textit{Action and Adventure}, \textit{1996-1999}), and aggregation functions (e.g., \textit{average}, \textit{sum}).
\snowy~uses a combination of statistical metrics derived from the underlying data and the interaction scores in the context state object to select the recommendation parameters.
Table~\ref{tab:parameter-selection} summarizes \snowy's logic for selecting parameters for different classes of intents.
The statistical functions in Table~\ref{tab:parameter-selection} are similar to those in prior work on insight- or data fact-based visualization recommendation systems (e.g.,~\cite{demiralp2017foresight,datasite,srinivasan2018augmenting}).

The parameter selection logic detailed in Table~\ref{tab:parameter-selection} is driven based on statistical metrics derived directly from the underlying data.
However, since the dataset remains constant throughout a session, if only these metrics were used to select parameters, the recommendations may get repetitive if users have already investigated a suggested combination of attributes.
To promote breadth in data exploration, besides ``data interestiness,'' \snowy~also incorporates prior interaction scores from the context map such that attributes and values with lower scores are bumped up when selecting parameters (\textbf{DG4}).

\begin{figure}[t!]
    \includegraphics[width=.6\linewidth,keepaspectratio]{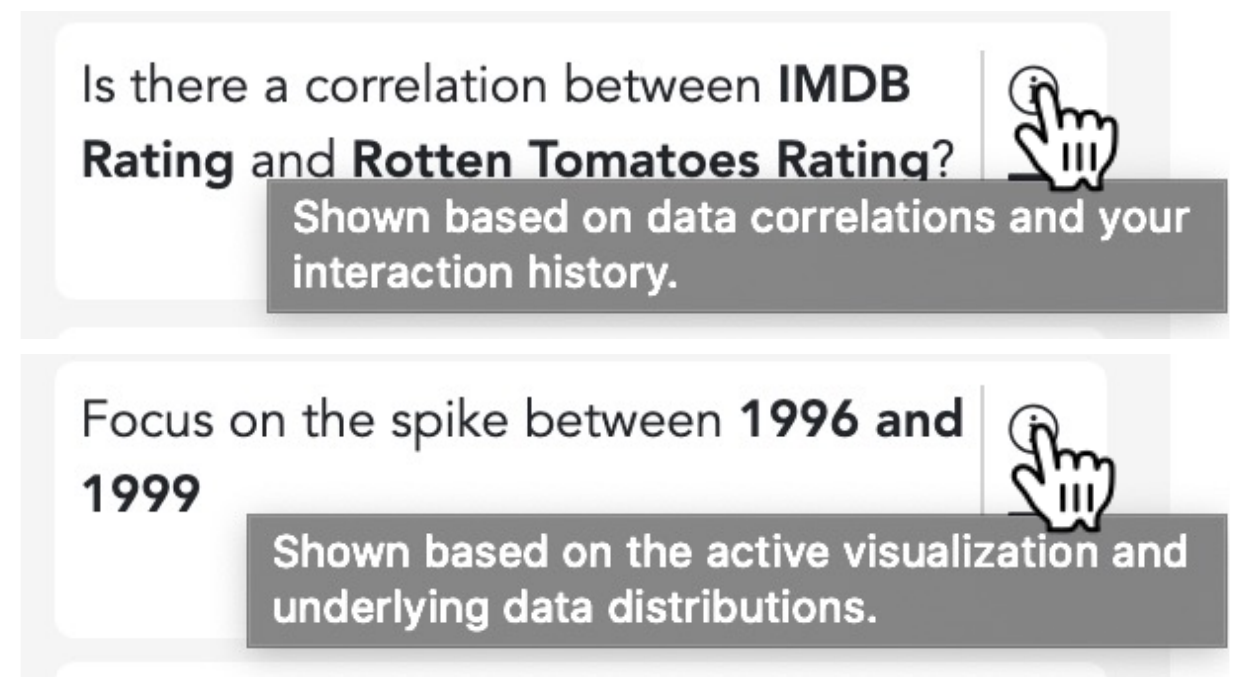}
    \caption{Examples of explanations for \intent{correlation} and \intent{filter} recommendations in \snowy's interface.}
    \label{fig:info-tooltips}
    \Description{Figure 8 displays examples of explanations accompanying recommendations in Snowy.}
\end{figure}

An example of the effect of this inclusion of interaction scores during parameterization can be seen in the aforementioned usage scenario by comparing the attributes included in the recommendations in Figures~\ref{fig:teaser}A,~\ref{fig:teaser}B,~\ref{fig:scenario-1} to those in Figures~\ref{fig:interface},~\ref{fig:scenario-2}.
Specifically, at the start of the session, the system recommends attributes like \textit{Content Rating}, \textit{Worldwide Gross}, and \textit{Major Genre} given the underlying data patterns.
However, as the session progresses, to promote data coverage, the recommendations shift to focus on attributes like \textit{Duration} and \textit{Creative Type} since Tintin has either never or only minimally investigated these attributes as part of his exploration (\textbf{DG1}).
Although \snowy~does not give users control over the parameter selection logic, it does provide a brief rationale for why recommendations are shown through tooltips in the interface (Figure~\ref{fig:info-tooltips}).

\begin{figure}[t!]
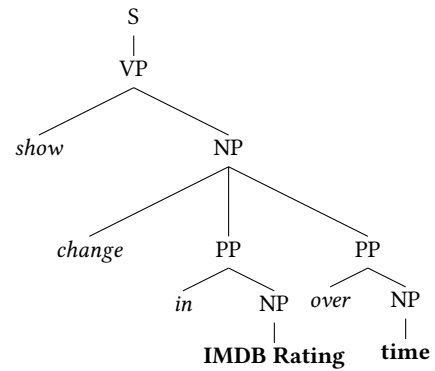

\Tree[.S   !\qsetw{.1cm} [.VP  \textit{show}
 !\qsetw{.2cm}
            !\qsetw{.1cm} [.NP \textit{change} 
               !\qsetw{.1cm}  [.PP \textit{in}
                !\qsetw{.2cm}     [.NP \textbf{IMDB Rating} ]]
              !\qsetw{1.8cm}    [.PP \textit{over}
               !\qsetw{1cm}       [.NP \textbf{time} ]]]]]
\caption{Syntax tree for the \intent{trend} intent that generates the recommendation, ``Show change in \texttt{IMDB Rating} over time''. \textbf{Terminals} are populated using the parameters passed to the \intent{trend} intent (the measure \textit{IMDB Rating} is applied as-is, whereas the dimension \textit{Release Year} is replaced with  ``time'').}
 \label{fig:parsetree}
 \Description{Figure 9 shows an example of a syntax tree for the recommendation ``Show change in IMDB Rating over time."}
\end{figure}

\subsubsection{Linguistic Realization of Recommendations}
\begin{table*}[t!]
\centering
\resizebox{.95\textwidth}{!}{%
\begin{tabular}{@{}ll@{}}
\toprule
\vspace{.5em}
Distribution &  \begin{tabular}{@{}ll@{}}\reco{Show the spread of values for \textbf{Rotten Tomatoes Rating}}, \reco{How many items exist for each \textbf{Creative Type}?}\\\reco{Count items by \textbf{Major Genre}}, \reco{What is the distribution of values for \textbf{Production Budget}?}\end{tabular} \\
\vspace{.5em}
Group        &  \begin{tabular}{@{}ll@{}}\reco{What is the average \textbf{Production Budget} by \textbf{Major Genre}?}, \reco{Compare average \textbf{Durations} across \textbf{Content Ratings}}\\\reco{On average, what is the \textbf{Production Budget} for each \textbf{Major Genre}?}\reco{Show total \textbf{Worldwide Gross} by \textbf{Creative Types}}\end{tabular}\\
\vspace{.5em}
Correlate    &  \begin{tabular}{@{}ll@{}}\reco{How does \textbf{IMDB Rating} vary with \textbf{Production Budget}?}, \reco{How are \textbf{Duration} and \textbf{IMDB Rating} correlated?}\\\reco{Show the relationship between \textbf{Worldwide Gross} and \textbf{Production Budget}}, \reco{How about \textbf{IMDB Rating} and \textbf{Rotten Tomatoes Rating}?}\end{tabular}\\
\vspace{.5em}Trend    &  \begin{tabular}{@{}ll@{}}\reco{Show the trend for \textbf{Worldwide Gross} instead}, \reco{What is the trend of \textbf{IMDB Rating} over the \textbf{years}?}\\\reco{How does \textbf{Production Budget} vary over \textbf{Release Years}?}, \reco{Plot changes in \textbf{Rotten Tomatoes Ratings} over \textbf{time}}\end{tabular}\\
\vspace{.5em}
Filter       &  \begin{tabular}{@{}ll@{}}\reco{Just show \textbf{Adventure, Action, and Musical}}, \reco{Focus on the spike between \textbf{1996} and \textbf{1999}},\\\reco{Just show the \textbf{top 3 groups}}, \reco{Focus on \textbf{high Production Budget}}, \reco{Drill down into \textbf{PG-13}}\end{tabular}\\
Aggregation & \reco{Show the \textbf{total Worldwide Gross} instead}, \reco{Show the \textbf{total} values instead}, \reco{How about the \textbf{mean of} values?}\\\bottomrule
\end{tabular}%
}
\vspace{.25em}
\caption{Examples of utterance recommendations generated by \snowy.}
\label{tab:nlg-examples}
\Description{Table 2 lists examples of utterance recommendations generated for the analytic intents supported by Snowy.}
\vspace{-2em}
\end{table*}


Once all the relevant parameters for the various recommendation types have been determined, they need to be combined together into a well-formed natural language utterance. This process called linguistic realization, involves ordering constituents of the recommendations and generating the right morphological forms (including verb conjugations and agreement)~\cite{reiterbook}. 
We employ a template-based approach for generating NL utterances as recommendations in \snowy. Given that the application domain is a set of known analytical intents along with attributes and values from the underlying datasets, the space of linguistic variations is relatively small and the outputs can be specified using templates~\cite{reiter:2010}.
\edited{Having a deterministic set of generated NL output also allowed us to control the variability in the recommended NL utterances for testing purposes.}
We defined the templates by referring to utterances commonly supported across existing NLIs~\cite{setlur2016eviza,hoque2017applying,yu2019flowsense,setlur2019inferencing,narechania2020nl4dv} and sample utterances collected through studies investigating the use of NL to create or interact with data visualizations~\cite{tory2019mean,srinivasan2021collecting}.
\edited{Note, however, that the current template-based approach can be extended to a task-oriented dialogue approach by using the set of templates along with a language model for generating a larger variety of sentences with linguistic variability.}

Our algorithm maps non-linguistic input comprising of data attributes, values, and intent from the parameterization process to a linguistic structure based on a set of predefined templates for each of the intents described in the \hyperref[sec:parameter-assignment]{previous section}. These templates contain gaps for the parameters and generate well-formed recommendation utterances when all the gaps have been replaced by linguistic structures that do not contain gaps. 

Formally, a template $T  = (S, E, C, I)$, where S is an abstract syntax tree (AST) for each analytical intent type $I$ with open slots in it; $E$ is a set of links to additional syntactic structures that are noun ($NP$), verb ($VP$), and prepositional phrases ($PP$) that are substituted in the gaps of $S$; $C$ is a set of analytical constraints on the applicability of $S$ that are based on the type of $I$. The interior nodes of $S$ are non-terminal symbols (i.e., syntactic variables) whose gaps are recursively replaced by groups of terminal symbols (i.e., elementary strings in the utterance) according to $T$'s production rules as well as synonyms and vocabulary terms based off of ~\cite{setlur2019inferencing}.

Now, consider the $S$ for the \intent{trend} intent with the parameters $measureStr$ and $dateTimeDimensionStr$. Figure~\ref{fig:parsetree} shows the AST for generating the recommendation, \reco{Show change in \textbf{IMDB Rating} } \reco{over \textbf{time}}. The parameters lead to generation of partial NPs, with the slots for $measureStr$ and $dateTimeDimensionStr$ filled by \textit{IMDB Rating} and \textit{time} respectively.
The linguistic realization process generates all possible valid ASTs from $T$, using a bottom-up generative process. Each AST in this set is checked to see whether it is compatible with $C$. After the set of utterances is generated, nouns and verbs are inflected, wherein the base forms of the words are modified to be grammatically sound to account for plurality and tense~\cite{crystal2011dictionary}. A variety of recommendation utterances are randomly generated to maximize the variety of utterance recommendations produced by \snowy.
Example utterance recommendations from the various \intent{trend} ASTs include \reco{What is the trend of \textbf{IMDB Rating} over the \textbf{years}?}, \reco{How does} \reco{\textbf{IMDB Rating} vary over \textbf{Release Years}?}, \reco{Show change in \textbf{IMDB Rat-} } \reco{\textbf{ing} over \textbf{time}}, and \reco{Show the change in \textbf{IMDB Rating} over the \textbf{years}}, among others.
From the resulting set of valid utterances, one is selected at random (in this case, \reco{Show change in \textbf{IMDB Rating} over \textbf{time}}).
Table~\ref{tab:nlg-examples} illustrates additional sample utterance recommendations for different intent types.

Notice from Table~\ref{tab:nlg-examples} that the recommendations exhibit a variety of linguistic variations to exemplify and help discover the capabilities of the underlying NL parser (\textbf{DG5}).
These variations include different phrasing patterns (e.g., questions, commands) and the use of colloquial terms in addition to the underlying data attributes and values (e.g., `over time' for temporal attributes, `spike' for a steep rise and fall trend in a line chart, and modifier terms like `low', `high' when referring to numeric filters).
Furthermore, some recommendations contain \emph{explicit} references to attributes, values, or intents and can be used as standalone utterances (e.g., \reco{What is the average \textbf{Production Budget} across \textbf{Major Genre}?}, \reco{Just s-} \reco{how \textbf{Adventure, Action, and Musical}}).
However, other recommendations include \emph{implicit} references to values and intents, and/or incorporate pragmatic markers to support conversational interaction (e.g., \reco{Just show \textbf{top 3} groups}, \reco{Now how about \textbf{IDMB Rating} and \textbf{Rott-}} \reco{\textbf{en Tomatoes Rating}?}).
The choice of which AST is used for generating the recommendation is based on the current context state and the analytical constraints $C$.
For example, in a case like Figure~\ref{fig:interface}, where the active chart is an unsorted bar chart of average \textit{Worldwide Gross} by \textit{Major Genre}, the filter recommendation is \reco{Just show \textbf{Adventure, Action, and Musical}} and explicitly lists the filter group.
However, in the case of Figure~\ref{fig:architecture}, when the context state is a sorted bar chart (since the invoking utterance ``Show \textit{highest} grossing genres'' includes an extremum token), the recommendation changes to \reco{Just show the \textbf{top 3} groups} as this phrasing suggests linguistic continuation and supports visual coherence~\cite{tory2019mean} by preserving the previous chart structure to show the top 3 groups in the sorted chart. As a session progresses, to make users aware of more advanced interpretation capabilities, \snowy~starts recommending utterances that combine intents (e.g., \reco{How has the \textbf{Production Budget}} \reco{changed over the \textbf{Release Years} for each \textbf{Creative Type}?} combining\\ \intent{trend} and \intent{group}, \reco{Show the relationship between \textbf{Rotten Tomatoes}} \reco{\textbf{Rating}  and \textbf{Duration} by \textbf{Major Genres}} combining \intent{correlation} and \intent{group}).

\section{Preliminary User Study}

We conducted a preliminary user study to gather initial feedback on the idea of presenting utterance recommendations during conversational visual analysis and assess the usability of the prototype.

\subsection{Participants and Setup}

We recruited $10$ participants (P1-P10, six males, four females) through a mailing list at a
data analytics software company. Participants were recruited on a first-come, first-serve basis. Based on self-reporting by the participants, five had never or infrequently performed data analysis, three occasionally performed data analysis, and two participants analyzed data on a daily basis. When asked about their prior experience level with interactive visualization tools like Tableau and Microsoft Power BI, four participants identified themselves as being expert users, three participants said they were familiar with the general capabilities of these tools and used them somewhat frequently, and three participants said they only occasionally used visualization tools.
Since the study involved NL interaction, we also asked participants about their prior experience level with NLIs for visualization including commercial systems like Tableau's Ask Data and Microsoft's Power BI Q\&A.
To this question, four participants said they had little to no experience using these tools, four participants said they frequently used the tools, and two participants said they seldom interacted with such tools, but were aware of their general capabilities. Participation in the study was voluntary and participants were not compensated for their time.

To conform with COVID-19 protocol, all sessions were conducted remotely via the Cisco WebEx video conferencing software~\cite{webex}. The prototype was hosted on a local server running on the experimenter's laptop\footnote{2.4 GHz MacBook Pro running macOS Catalina 10.15.7 set to a resolution of 3072 $\times$ 1920.}. Participants were granted control over the experimenter's screen during the session (the setup was tested through three pilot studies to ensure there was no lag or technical issues in the interaction experience). All studies followed a think-aloud protocol.
The audio, video, and on-screen actions were recorded for all sessions with permission from the participants.

\subsection{Procedure}

Sessions lasted 42--53 minutes (mean: 49 min.) and were roughly organized as follows:

\vspace{.5em}
\noindent [0--10 min.]: Introduction to the study goals and time for participants to fill out their background information. Participants were briefly introduced to \snowy's interface. Since an implicit goal was to assess if the recommendations could help with NL input, the introduction for the NL and recommendation components of the interface were kept to a bare minimum (e.g.~how to execute NL utterances or select recommendations) to avoid participant bias.

\vspace{.5em}
\noindent [10--25 min.]: Participants were given a set of five tasks involving the movies dataset introduced earlier in Section~\ref{sec:interface-and-scenario} and were asked to ``solve'' them using \snowy.
These tasks involved a combination of directed exploration tasks where participants were asked to explore the data with respect to a subset of attributes (e.g.~``List 1--3 insights pertaining to the Content Rating attribute'') and Jeopardy-style fact verification tasks, similar to those used in~\cite{datatone} where participants were given a fact and had to ask questions of the data to determine if the fact was either true or false. The tasks were framed such that directly typing the instructions into the system would not result in the answer. The order of tasks was randomized across participants.

\vspace{.5em}
\noindent [25--40 min.]: Participants were then given a second dataset about 500 colleges in the U.S.
and were asked to freely explore it with \snowy.
The dataset contained nine attributes for each college including three categorical attributes ({\small{\faFont}}~\textit{Region}, {\small{\faFont}}~\textit{Locale}, {\small{\faFont}}~\textit{Control}) and six numerical attributes (e.g., {\small{\faHashtag}}~\textit{Admission Rate}, {\small{\faHashtag}}~\textit{Cost}, {\small{\faHashtag}}~\textit{Debt}).
Incorporating both targeted and open-ended exploration allowed us to assess the impact of utterance recommendations across the two popular scenarios for data analysis and validate our specific design goals \textbf{(DG1, DG6)}.

\begin{figure}[t!]
    \includegraphics[width=.7\linewidth,keepaspectratio]{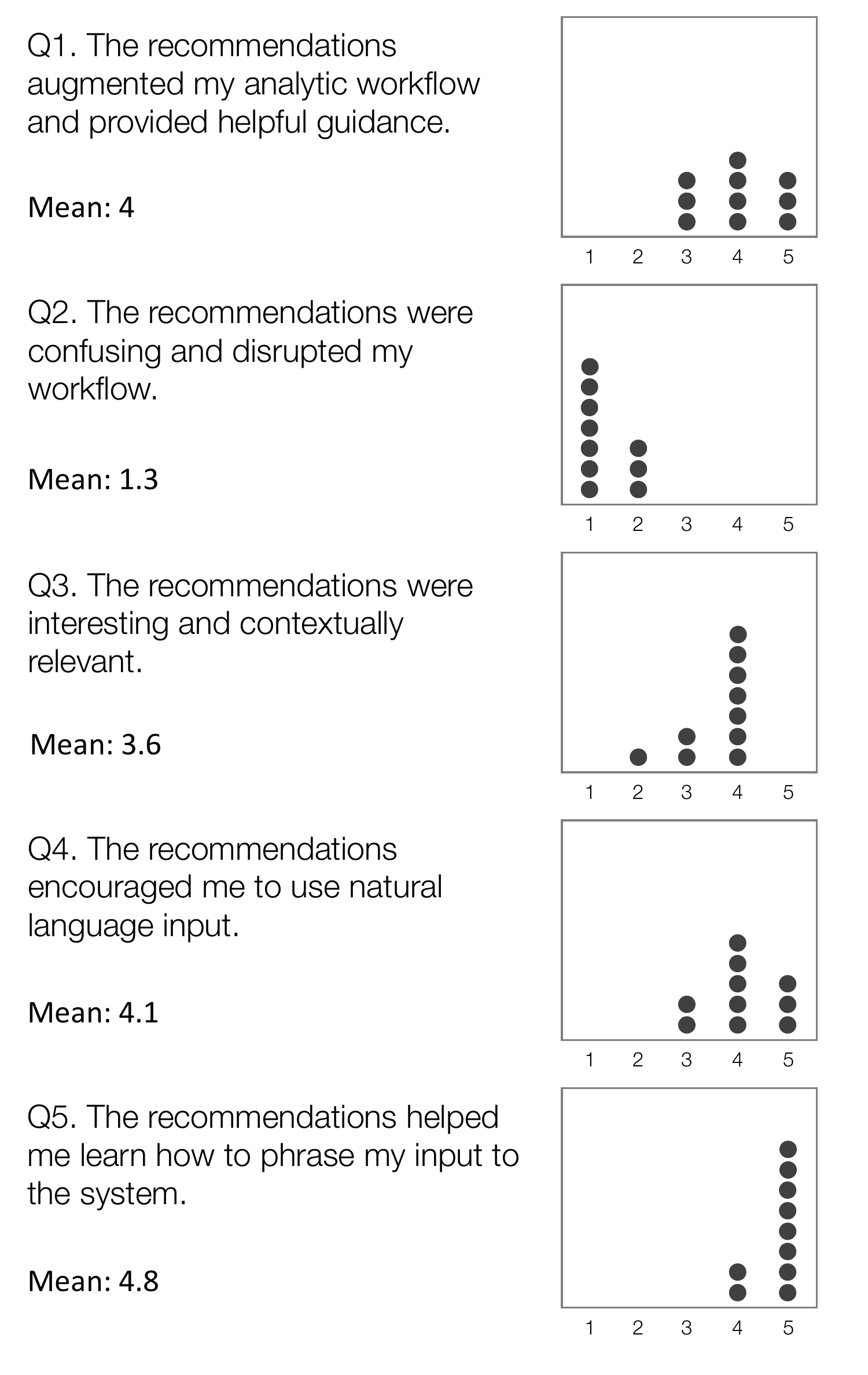}
    \caption{Participant responses to post-session questions about utterance recommendations in \snowy. Statements were rated on a scale of 1 (Strongly Disagree) to 5 (Strongly Agree).}
    \label{fig:results}
    \Description{Figure 10 summarizes participant responses to post-session questions. Participants were generally in agreement about recommendations being helpful guidance tools during visual analysis but there were mixed responses about the relevance of the current prototype's recommendations. However, all participants strongly agreed that the recommendations encouraged them to use natural language input and helped them learn how to interact with the system through language.}
\end{figure}

\vspace{.5em}
\noindent [40--50 min.]: Post-session questionnaire on \snowy's recommendations (Figure~\ref{fig:results}) along with ten questions from the standard System Usability Scale (SUS) questionnaire~\cite{sus} to help evaluate the prototype's usability. The questionnaire was complemented with a semi-structured interview where participants talked about their experience using \snowy.

The experimenter script, task descriptions, and datasets are included in supplementary material. 

\subsection{Results and Discussion}
On average, participants completed four out of the five tasks in the first targeted-exploration phase, spending between 12--18 min. (mean: 14 min.), followed by 8--17 min. (mean: 14 min.) on open-ended exploration with the college data.
Participants gave \snowy~an average SUS score of
$76.5$ (a score of $\ge 68$ is considered as an indicator of good usability~\cite{sus}).

\subsubsection{Recommendations for guiding visual analysis.}

Participants generally agreed that the recommendations supported their analytic workflows and provided useful guidance during data exploration (Figure~\ref{fig:results}, Q1).
For instance, P4 saw the recommendations as good starting points during data exploration and said, ``\textit{I was curious about what \snowy~was picking up and just kind of clicking through those different recommendations to see what the data looks like.}''
P2 noted that the ``\textit{recommendations were pretty well spread out}'' and helped him see the data from different perspectives (i.e., supported different intents, \textbf{DG3}).
Recommendations also reminded participants of the analytical capabilities of the tool and helped pivot their exploration around those.
For instance, while looking at a bar chart showing total values, P4 saw a recommendation about the \textit{average} values and realized he could switch aggregations to gain different insights about the same attribute combinations.
All participants stated that they found the recommendations more useful during the open-ended exploration.
For instance, P9 said ``\textit{it was really helpful for exploratory, but not so much in the first focused part. For the last few tasks I didn't even look at the recommendations...}''
This reaction was not surprising as \snowy~was designed with data exploration in mind (as opposed to targeted analysis).
That said, participants also confirmed that the recommendations did not obstruct their workflow during targeted analysis (\textbf{DG6}) and they could ``\textit{simply not look}'' when they knew what actions they wanted to perform (Figure~\ref{fig:results}, Q2).

\subsubsection{Relevance of recommended content.}
\label{sec:results-feedback-relevance}
Although participants were in agreement about the utility of recommendations to guide their exploration, the feedback on the relevance of recommended utterance parameters (i.e., attributes and values) was rather mixed (Figure~\ref{fig:results}, Q3).
Some participants felt that the recommendations did a good job at suggesting what they wanted to explore next. P3 cited an example from her session and said, ``\textit{I had some assumptions in my head, like, for the movies, maybe the budget would be somehow related to, you know, Worldwide Gross and it is quite delightful when I found that was actually a recommended query too.}"
In other cases, especially sessions involving participants with prior experience with NLIs, we observed various assumptions that participants made as they drew from their past experiences.
For instance, these participants assumed that the recommendations were randomly parameterized and thus only viewed them as sample phrases.
For instance, P8 said, ``\textit{the odds that you're going to guess the fields I actually want to look at are terribly low. So I was looking at what is the verbiage around the fields trying to pull that out}.''
Similarly, P1 said, ``\textit{I'm not really understanding why it gave the choices. Like, is it's based on previously asked questions or what? So I thought they were relevant as, like, oh, let me explore, but I'm not exactly sure}.''
While these specific participants later noted that they had forgotten about the tooltips that provide high-level explanations for recommendations, their feedback suggests the need for surfacing better explanations in the interface.

\subsubsection{Recommendations to aid NL discovery and usage.}

All participants were in strong agreement that the recommendations helped them discover and learn the various ways NL could be used as input to the system (Q4 and Q5 in Figure~\ref{fig:results}).
For instance, P2 said, ``\textit{the most valuable part of the recommendations was to know how to describe to the system, the kinds of things I'd want to see. It's a really good tutorial.}''
P5 highlighted how the recommendations were useful as phrasing templates even when they did not suggest the exact content that she was looking for - ``\textit{Obviously you can't read my mind. So, when I was looking for something very specific, it wouldn't exactly be there. But it would be a good be guiding tool to know how I could ask questions.}''
Participants also appreciated the linguistic variety in the recommendations to help them understand the different ways that the NL utterances could be expressed.
P9, for instance, said, ``\textit{seeing the recommendations helped me because I could see, like, oh, I can phrase it as a question or I can use synonyms or, you know, that gave me information I needed and seeing examples is really helpful.}''
Commenting specifically on the follow-up recommendations, P10 said ``\textit{Words like `drill down' are helpful to indicate that it's the same inquiry but just like a follow-up on it. And words like `instead' also helped}.''

\vspace{.5em}
\noindent{}The individual participant responses (Figure~\ref{fig:results}) and the subjective feedback together helped us validate our high-level premise that utterance recommendations can guide visual analysis while implicitly making users aware of the system's NL capabilities. Participant feedback suggested that while the current recommendations in \snowy~provide good linguistic variety, there is a need for providing users better explanations about the provenance of the system's behavior.

\subsection{Design Considerations}

Findings from the study brought up three important factors to consider when incorporating utterance recommendations in the context of conversational visual analysis tools:
\begin{itemize}[leftmargin=.2in]
    \item \textbf{Show visual previews of system changes.} Participants who were familiar with visualization tools and concepts like graphical encodings commented that it would be helpful if the system provided additional feedback or even a preview for what actions selecting a recommendation would result in. For instance, referring to recommendations like \reco{Break down by \textbf{Content Rating}}, P4 said, ``\textit{I got the sense of what it's recommending, but at the same time, I wasn't sure what that's going to do to the visualization.}'' Specifically, in this case, P4 was unsure if ``break down'' would lead to \textit{Content Rating} being mapped to color or if the system will somehow create a small multiples chart instead.
    \item \textbf{Adaptable interfaces for managing recommendations.}
    Following \textbf{DG1} and \textbf{DG4}, we designed the recommendations in \snowy~to be \textit{adaptive} to the user's interaction based on available data entities (e.g., underexplored attributes are prioritized during parameterization, recommendations mapping to infrequently used intents are shown first).
    While this functionality aids breadth-oriented exploration, it makes it challenging to go back to prior recommendations as their ordering or parameters might have changed~\cite{findlater2009design}.
    For example, P10 had selected a recommendation and wanted to go back to a recommendation she saw earlier, but did not want to investigate at the time (``\textit{I saw a suggestion here before I clicked some other suggestions.. Is there any way to request the same examples?}'').
 Addressing such scenarios and exploring alternative \textit{adaptable interfaces}~\cite{stuerzlinger2006user,findlater2008impact} that allow users to bookmark, organize, and track utterance recommendations over time, would be an interesting approach to pursue.
    \item \textbf{Placement of recommendations.} Recommendations are persistently displayed on the right side panel of the interface (Figure~\ref{fig:interface}E). Participants generally liked this placement for convenient reference, while not impeding their workflow during targeted analysis (\textbf{DG6}). For follow-up recommendations however, we observed that participants sometimes missed noticing the recommendations although they were suggesting the exact intent and attribute(s) they were interested in (e.g.~filtering or grouping a bar chart by a specific categorical attribute like \textit{Region}). To this end, P3 suggested that an alternative would be to show the recommendations, especially follow-ups, closer to the the input box since they are typically the most immediate actions one may want to take.
    However, doing so could be distracting while users are typing their questions, especially if the recommendations frequently update. Thoughtful placement of recommendations in such an interface would need to strike a balance between non-obtrusiveness and in-situ availability.
\end{itemize}

\section{Limitations and Future Work}

\noindent{}\textbf{Supporting additional visualizations and intent categories.}
The current encoding channels of \texttt{x}, \texttt{y}, and \texttt{color} along with the \texttt{bar}, \texttt{point}, and \texttt{line} marks allow the creation of canonical visualization types and provide enough variability to test the underlying idea of leveraging utterance recommendations.
However, moving forward, for the system to have practical value, more data types and visualizations (e.g., maps, small multiples) need to be supported.
Furthermore, while our focus in this paper was on core visual analysis intents, people may want to use NL to accomplish other tasks like styling a visualization or for user interface operations at the tool level. For instance, during the study, two participants (P2, P8) said they would like the utterances to cover a more general set of actions and have a tighter coupling with the GUI (e.g., removing attributes through NL, changing the colors used in a chart). While supporting these, in theory, can be done by adding more types of recommendations (e.g., analytic intents, visualization styling), thinking about the manifestation of these recommendations in the interface and the ways to delineate different recommendation types are open points for future work.

\vspace{.5em}
\noindent{}\textbf{\edited{Investigating the impact of utterance recommendations on analytic workflows and learnability.}}
\edited{The preliminary study helped us validate the premise that utterance recommendations can guide visual analysis while aiding with NL discoverability. However, deeper investigations are required to understand the specific benefits of utterance recommendations and their impact on analytic workflows. Future work could involve running a study comparing \snowy~to a baseline tool that recommends visualizations (as opposed to NL utterances). Such a study could help better understand the unique pros and cons of utterance recommendations and shed light on design considerations to combine utterance recommendations with thumbnail-style chart previews. Participants positively commented on the potential of utterance recommendations to aid NL discovery and learning, with P7 comparing \snowy's utterances to command suggestions offered by voice assistants like Amazon Alexa and Google Home. However, this feedback was based on a single session involving $\sim$30 minutes of interaction with the tool. To truly assess the usefulness of NL recommendations, it would be necessary to conduct a longitudinal study with \snowy~where participants use the interface with their individual datasets and workflows over a longer time period. 
}

\vspace{.5em}
\noindent{}\textbf{\edited{Mitigating potential biases.}}
\edited{To promote breadth of coverage in its recommendations, \snowy's recommendation engine tracks and incorporates the count of interactions with attributes and intents.
`Coverage', however, is a very simplistic metric and may not account for the various types of analytic and cognitive biases~\cite{wall2018four,dimara2018task}, or prevent people ignoring their external knowledge about the data domain and blindly follow the recommendations.
It is critical to incorporate richer models to identify potential biases in user interactions while generating the recommendations (e.g.~by leveraging cognitive bias metrics such as those suggested by Wall et al.~\cite{wall2017warning}) and investigate designs to surface and mitigate potential biases during exploratory data analysis~\cite{wall2019toward}.
}

\vspace{.5em}
\noindent{}\textbf{Applying machine learning approaches for utterance realization.}
While the current template-based approach for utterance realization works effectively for a small set of known intents, it can be rather challenging to configure templates for large-scale systems that cover a more comprehensive range of intents and data domains. An interesting direction to pursue is applying machine learning for supporting linguistic realization. Recent deep-learning language models such as GPT-3~\cite{brown2020language} could also be an alternative to consider for providing greater linguistic variability in utterances at low configuration cost.

\vspace{.5em}
\noindent{}\textbf{Incorporating data semantics during utterance realization.}
While inferring the semantics of the underlying data to generate domain-specific phrasings and recommendations is an open, vast area for research, we also identified more short-term ways to improve the recommendations by incorporating data semantics.
For example, one of the participants (P7) suggested that if the recommendations were rephrased to use terms from the data domain (e.g., ``movies'' or ``colleges'' instead of ``items''), they would be more engaging and interpretable, especially for non-technical domain-experts.
Improving the language of the recommendations and investigating such simple, but important semantic modifications and data curation is another area for improvement going forward.

\vspace{.5em}
\noindent{}\textbf{Exploring voice input and chatbots.}
In this paper, we focus on \snowy~as a desktop system that supports NL interaction through text input. The premise of leveraging utterance recommendations for conversational visual analysis is generalizable and can be applied in other applications and contexts.
For instance, prior work on multimodal interfaces for data visualization involving voice input (e.g.,~\cite{saktheeswaran2020touch,srinivasan2020interweaving}) has highlighted that NL discovery is a persistent challenge. Along these lines, a compelling opportunity for future research lies in exploring how utterance recommendations similar to those in \snowy~can be generated and surfaced in the context of voice-based interfaces to data (as opposed to text). Exploring these alternative interfaces would need to consider user context, device modalities, as well as differences in language pragmatics and syntax structure for generating recommendations.

\section{Conclusion}
NLIs for visual analysis tools have evolved as a promising medium for users to converse with data and gain insights by expressing their inquiries in simple language. The process of sense-making and getting insights from the data continues to be a challenge as users need to formulate their questions while making progress in their analytical journey. In this paper, we introduce a mixed-initiative system, \snowy~that provides \emph{both} analytical and linguistic guidance to the user by presenting utterance recommendations. The system suggests new inquiries as well as follow-up utterances based on the user's current context, providing useful next steps for interesting and underexplored aspects of the data. A preliminary evaluation of \snowy~suggests that contextual utterance recommendations can not only guide visual analysis, but also help people gain awareness of the system’s NL interpretation capabilities.  We hope that insights learned from our work can inspire new research directions in the combined space of NL, recommendations, and analytical capabilities. As we move a step closer towards realizing richer analytical conversation experiences during visual analysis, the excitement for potential innovation and opportunity can be best expressed by Captain Haddock~\cite{haddock}, a fictional character in The Adventures of Tintin as he exclaims - ``Blistering blue barnacles!''


\begin{acks}
We thank Clark Wildenradt for his input and suggestions for improving \snowy's interface. We also thank members of the Tableau Research Team and the anonymous reviewers for their helpful feedback, and all the study participants for their opinions and insights.
\end{acks}

\bibliographystyle{ACM-Reference-Format}
\bibliography{references}

\end{document}